\documentclass[12pt,draftclsnofoot,onecolumn]{IEEEtran}

\usepackage{cite}
\ifCLASSINFOpdf
  \usepackage[pdftex]{graphicx}
\else
  \usepackage[dvips]{graphicx}
\fi

\pdfoutput=1

\usepackage[cmex10]{amsmath}
\usepackage{amsmath,amsfonts,amssymb}
\usepackage{mdwmath}

\usepackage{algorithmic}

\usepackage[ruled,lined,boxed]{algorithm2e}

\usepackage{array}
\usepackage{fixltx2e}
\usepackage{bm}
\usepackage{stfloats}
\usepackage{graphicx}
\usepackage{caption}
\usepackage{mdwtab}
\usepackage{subfig}
\usepackage{booktabs}
\usepackage{multirow}
\usepackage{etoolbox}
\usepackage{makecell}
\usepackage{xcolor}
\usepackage[bottom]{footmisc}
\usepackage{mdframed}
\usepackage{steinmetz}

\usepackage{setspace}

\newcommand{\e}{\begin{equation}}
\newcommand{\ee}{\end{equation}}
\newcommand{\eqn}{\begin{eqnarray}}
\newcommand{\eeqn}{\end{eqnarray}}

\begin{document}
\title{Data-Driven Deep Learning Based Hybrid Beamforming for Aerial Massive MIMO-OFDM Systems with Implicit CSI}

\author{Zhen~Gao, Minghui~Wu, Chun~Hu, Feifei~Gao,  Guanghui~Wen, Dezhi~Zheng, Jun~Zhang
\vspace*{-5.0mm}
\thanks{Z.~Gao, M.~Wu, C.~Hu, D.~Zheng, and J.~Zhang are with School of Information and Electronics, Beijing Institute of Technology, Beijing 100081, China (E-mails: gaozhen16@bit.edu.cn; wuminghui@bit.edu.cn; bit\_hc@bit.edu.cn; zhengdezhi@bit.edu.cn; buaazhangjun@vip.sina.com).}
\thanks{F. Gao is with the Institute for Artificial Intelligence, Tsinghua University (THUAI), State Key Lab of Intelligent Technologier
	and Systems, Tsinghua University, Beijing National Research Center for Information Science and Technology (BNRist), Department
	of Automation, Tsinghua University, Beijing 100084, China (E-mail: feifeigao@ieee.org).}
\thanks{G. Wen is with the Department of Mathematics, Southeast University, Nanjing 210096, China   (E-mail: wenguanghui@gmail.com).}
}

\begin{spacing}{1.21}
\maketitle

\begin{abstract}
In an aerial hybrid massive multiple-input multiple-output (MIMO) and orthogonal frequency division multiplexing (OFDM) system, how to design a spectral-efficient broadband multi-user hybrid beamforming with a limited pilot and feedback overhead is challenging. 
To this end, by modeling the key transmission modules as an end-to-end (E2E) neural network, this paper proposes a data-driven deep learning (DL)-based unified hybrid beamforming framework for both the time division duplex (TDD) and frequency division duplex (FDD) systems with implicit channel state information (CSI). 
For TDD systems, the proposed DL-based approach jointly models the uplink pilot combining and downlink hybrid beamforming modules as an E2E neural network. While for FDD systems, we jointly model  the downlink pilot transmission, uplink CSI feedback, and downlink hybrid beamforming modules as an E2E neural network. Different from conventional approaches separately processing different modules, the proposed solution simultaneously optimizes all modules with the sum rate as the optimization object.
Therefore, by perceiving the inherent property of air-to-ground massive MIMO-OFDM channel samples, the DL-based E2E neural network can establish the mapping function from the channel to the beamformer, so that the explicit channel reconstruction can be avoided with reduced pilot and feedback overhead. Besides, practical low-resolution phase shifters (PSs) introduce the quantization constraint, leading to the intractable gradient backpropagation when training the neural network. To mitigate the performance loss caused by the phase quantization error, we adopt the transfer learning strategy to further fine-tune the E2E neural network based on a pre-trained network that assumes the ideal infinite-resolution PSs.
Numerical results show that our DL-based schemes have considerable advantages over state-of-the-art schemes. 
\end{abstract}


\IEEEpeerreviewmaketitle

\end{spacing}
\begin{spacing}{1.32}
\section{Introduction}\label{S1}


The space-air-ground-sea integrated network (SAGSIN) has been widely recognized as a promising architecture to enable the future sixth-generation (6G) communications paradigm shift for seamless global coverage \cite{UAV_xiao1}. In SAGSIN, aerial networks consisting of unmanned aerial vehicles, airships, and balloons based aerial base stations (BSs) are key components to support the wide-area coverage for scenarios not well covered by conventional terrestrial networks, such as cell-edge user service, agricultural monitoring, emergency communication, disaster rescue, remote areas, etc \cite{UAV_xiao3}. Communication capacity is crucial for aerial networks to support these application scenarios. To this end, the aerial BSs can be equipped with massive multiple-input multiple-output (MIMO) to substantially improve the system capacity \cite{UAV_hybrid}. However, the deployment of massive antenna arrays at the aerial BSs leads to the huge dimensions of channel state information (CSI), which bring difficulties to the real-time CSI acquisition and beamforming. Therefore, developing efficient physical layer techniques to meet the SAGSIN's high throughput and low latency requirements is urgent.

In time division duplex (TDD) systems, the BS with fully-digital array can easily obtain the uplink CSI based on the uplink pilot transmitted by the user equipments (UEs), and then exploits the channel reciprocity to obtain the downlink CSI for beamforming. However, aerial BS, e.g., airships and balloons, can not afford fully-digital array, due to its prohibitive power consumption and hardware cost \cite{UAV_xiao2}. 
Hybrid analog-digital array deployed at the aerial BS can well overcome the above issues \cite{UAV_hybrid}, but estimating the high-dimensional channels from a limited number of radio frequency (RF) chains at the aerial BS still suffers from high pilot overhead \cite{TDDHard}. 
Moreover, for frequency division duplex (FDD) systems \footnote{Compared with TDD systems, FDD systems have lower transmission delay and larger coverage radius. Therefore, when aerial BS is used for hotspot coverage with relatively low altitude, TDD is prioritized; while for wide area coverage with relatively high altitude, FDD is prioritized.}, the uplink and downlink channel’s reciprocity does not exist, leading the downlink CSI acquisition more challenging. Specifically, obtaining the downlink CSI in FDD systems relies on the downlink pilot transmitted by the BS, based on which the received pilot signals or the estimated channels are first compressed and quantized at the UEs, and then fed back to the BS for downlink beamforming. Due to the hundreds of antennas at the aerial BS, the resulting high-dimensional downlink CSI matrix would pose a prohibitively high overhead to downlink channel estimation and uplink CSI feedback at the UEs \cite{ins5}. 
Additionally, even in the case of perfect downlink CSI at the aerial BS, designing spectrum-efficient multi-user broadband hybrid beamforming is intractable, since the RF analog beamforming part is frequency-flat while the practical broadband channels are frequency-selective \cite{hybrid_flat}.
 Therefore, how to achieve accurate CSI at the aerial BS and perform reliable multi-user hybrid beamforming is a critical issue.

 \subsection{Related Works}\label{S1.1}
 
As for the channel estimation in TDD hybrid MIMO systems,
 the sparsity of millimeter wave (mmWave) channels in the delay or angle domain can be exploited to reduce the pilot overhead with the aid of compressed sensing (CS) techniques \cite{CS}. 
 The authors in \cite{greedy1,greedy2,greedy3,greedy4} adopt the greedy CS algorithms including orthogonal matching pursuit (OMP) and simultaneous weighted (SW)-OMP for CSI acquisition, while the authors in \cite{bay1,bay2,bay3} adopt the bayesian CS algorithms, such as approximate message passing (AMP) algorithm.  However, these channel estimation methods highly rely on \emph{a priori} information of the channels, which would inevitably lead to the performance loss when the assumed \emph{a priori} information is inconsistent with the practical channels. 

As for the channel estimation and channel feedback in FDD massive MIMO systems, 
one of the existing schemes is to exploit the sparsity of channels in the angle domain and delay domain to reduce the pilot and feedback overhead \cite{ins5,ins6,ins7}. The UEs can obtain the sparse channel parameters based on the received downlink pilot signals with the aid of CS techniques, and then quantize these parameters and feed them back to the BS for channel reconstruction \cite{b3}. Some other existing conventional schemes are based on codebooks, such as discrete Fourier transform (DFT) codebook \cite{codebook}. 
Specifically, the BS and UEs perform beam searching according to the predefined codebooks, and the UEs will feedback the indices 
of multiple beam pairs with relatively strong received signal-to-noise ratios (SNRs) \cite{codebook,ins8}. In \cite{Linear}, the comparison between these two methods has proved that the first kind of methods that feed the sparse channel parameters back to the BS is more efficient when the feedback overhead is limited.

As for the hybrid beamforming, a hierarchical codebook-based hybrid beamforming scheme was proposed in \cite{UAV_xiao5} for single-user MIMO systems.
By exploiting the channel sparsity, spatially sparse hybrid beamforming (SS-HB) and its derivatives were respectively proposed in \cite{OMP_HP1,OMP_HP2} to achieve enhanced performance. 
{\color{black}In \cite{Alt1} and \cite{Alt2}, the authors proposed alternative minimization hybrid beamforming schemes to approach the performance of the fully digital beamformer.}
Moreover, a two-stage hybrid beamforming (TS-HB) scheme and a heuristic hybrid beamforming scheme were respectively proposed in \cite{TSHP} and \cite{heuristic_HP} for multi-user MIMO systems. 
{\color{black}For the case where  the number of users is greater than the number of RF chains, the authors of \cite{NOMA1} and \cite{NOMA2} proposed hybrid beamforming schemes based on non-orthogonal multiple access.
In \cite{low1}, the authors proposed a hybrid beamforming scheme with low-resolution phase shifters (PSs).}
Due to the frequency-selective-fading channels in practice, orthogonal frequency division multiplexing (OFDM) is widely adopted to combat the multipath effect in broadband systems.
Therefore, the broadband hybrid beamforming schemes based on alternative minimization \cite{xiang_JSTISP}, principal component analysis (PCA) \cite{Sun_TWC}, array steering vector codebook \cite{codebook_HP}, {\color{black}and weighted sum rate maximization \cite{WMMSE1} were respectively proposed for MIMO-OFDM systems.} However, these hybrid beamforming schemes need either perfect downlink CSI or a codebook with an accurate sparse basis, which are difficult to acquire in practical systems.

In recent years, deep learning methods have been successfully applied in various fields, and also widely studied in the communication area, including CSI feedback \cite{csinet,quancsinet,GF_CSI}, channel estimation \cite{Ma_tvt,CE1_JSAC,CE2_JSAC}, beamforming \cite{DLanalog,DL_unfloding,Yuwei,E2EHB,E2EHB2,E2EHB3}, and signal detection \cite{SD1_JSAC,SD2_JSAC}, etc. Specifically, in terms of CSI feedback, a convolutional neural network (CNN)-based autoencoder named csiNet was proposed in \cite{csinet} to efficiently compress and accurately recover CSI. Besides, \cite{quancsinet} further investigated the bit quantization of CSI feedback and made some improvements to csiNet. Due to the interference and nonlinear factors in practical CSI feedback, a deep learning-based denoising network was proposed in \cite{GF_CSI} to improve the performance of CSI feedback. 
In terms of channel estimation, the authors of \cite{Ma_tvt} proposed a CNN-based channel estimation scheme for hybrid massive MIMO systems, while the authors of \cite{CE1_JSAC} proposed a model-driven DL-based solution.
{\color{black}In terms of beamforming, the authors of \cite{DLanalog,DL_unfloding,DL_HB1} proposed the DL-based analog beamforming and hybrid beamforming schemes. 
However, these DL-based beamforming schemes require perfect explicit CSI, which can be difficult to obtain in FDD massive MIMO with fully-digital array and TDD/FDD massive MIMO with hybrid array. 
To this end, some DL-based beamforming schemes that only require implicit CSI were proposed to avoid explicit channel acquisition \cite{Yuwei,E2EHB,E2EHB4,E2EHB2,E2EHB3}.
Specifically, the authors of \cite{Yuwei} jointly modeled the pilot design, channel feedback, and digital beamforming as an end-to-end (E2E) neural network for narrowband FDD fully-digital MIMO systems. 
As for DL-based hybrid beamforming, the authors of \cite{E2EHB,E2EHB4,E2EHB2} proposed joint channel sensing and hybrid beamforming schemes based on DL for TDD MIMO systems, in which \cite{E2EHB2} considers the quantization of PSs' phase values.
Furthermore, the authors of \cite{E2EHB3} proposed a DL scheme based on quantized received signal strength indicators to design hybrid beamforming for FDD MIMO systems. 
However, these DL-based hybrid beamforming schemes with implicit CSI either only focus on the analog channel sensing and beamforming design in TDD systems without considering the joint design of analog and digital parts \cite{E2EHB,E2EHB4,E2EHB2}, or only focus on the hybrid beamforming design in FDD systems without considering the design of pilot and CSI feedback \cite{E2EHB3}. In general, the existing DL-based schemes can efficiently improve the system performance through data-driven training, but there is still no unified E2E DL framework for both TDD and FDD multi-user broadband hybrid massive MIMO-OFDM systems.
}


%
\subsection{Motivations}\label{S1.2}

As for the conventional model-driven schemes \cite{greedy1,greedy2,greedy3,greedy4,bay1,bay2,bay3,ins5,ins6,ins7,b3,codebook,ins8,Linear,OMP_HP1,OMP_HP2,UAV_xiao5,TSHP,heuristic_HP,xiang_JSTISP,Sun_TWC,codebook_HP,NOMA1,NOMA2,low1,Alt1,Alt2}, pilot design, channel estimation, CSI feedback,
and hybrid beamforming are regarded as mutually independent modules to be separately optimized. Therefore, these schemes usually suffer from prohibitively high CSI estimation and feedback overhead for aerial hybrid massive MIMO-OFDM systems. Meanwhile, the inconsistency between the ideal \emph{a priori} information assumed by model-based schemes and the practical imperfect factors would further degrade the performance. 
By contrast, the data-driven schemes can learn a wealth of \emph{a priori} CSI through training samples, which avoids the dependence of traditional schemes on \emph{a priori} CSI assumptions and reduces the pilot and feedback overhead \cite{csinet,quancsinet,GF_CSI,Yuwei,E2EHB,E2EHB4,E2EHB2,E2EHB3}.
 Moreover, recent research in \cite{Yuwei} has demonstrated that the joint optimization of different signal processing modules can achieve better performance.   
Therefore, for aerial hybrid massive MIMO-OFDM systems, we propose to leverage the data-driven deep learning (DL) signal processing paradigm to jointly optimize the pilot design, channel estimation, CSI feedback, and hybrid beamforming for both TDD and FDD systems.  
On the other hand, since the intractable gradient backpropagation  issue would be caused by the quantization constraint of the low-resolution PSs, the aforementioned DL-based analog beamforming schemes \cite{DLanalog,DL_unfloding,DL_HB1,E2EHB,E2EHB4,E2EHB3} assuming the ideal infinite-resolution PSs in training stage
 will inevitably suffer from the performance loss in the practical systems with low-resolution PSs. 
 {\color{black} Besides, \cite{E2EHB2} uses a tanh-like function to approximate the PSs' phase values quantization in training stage, and uses an ideal quantizer in the test stage, which will still lead to the mismatch between training stage and test stage.}
Therefore, how to reduce the performance loss caused by phase
quantization error needs to be further investigated.

\subsection{Contributions}\label{S1.3}
In this paper, we develop a DL-based E2E joint training method for aerial massive MIMO-OFDM systems with hybrid beamforming.
Through data-driven training, the proposed E2E neural networks can obtain \emph{a priori} CSI from training samples, thus reducing the pilot and feedback overhead. 
By modeling the key transmission modules as an E2E neural network, we jointly optimize all modules with the goal of maximizing spectral efficiency.
Our main contributions can be summarized as follows:

\begin{itemize}
	\item {\color{black}We propose an E2E DL framework for TDD massive MIMO-OFDM hybrid beamforming systems, where the key transmission modules including uplink pilot combination and downlink multi-user hybrid precoding modules are modeled as an E2E neural network. By taking the negative spectral efficiency as the loss function and performing data-driven E2E training, we realize the joint optimization of these two modules, avoid the  explicit channel estimation with reduced pilot overhead, and achieve better performance over state-of-the-art schemes in TDD massive MIMO-OFDM hybrid beamforming systems.}
\end{itemize}
\begin{itemize}
	\item We extend the proposed DL-based E2E scheme from TDD mode to FDD mode, which models the key transmission modules including downlink pilot transmission, uplink CSI feedback, and downlink hybrid beamforming modules as an E2E neural network. {\color{black}By using data-driven E2E training, the proposed scheme can avoid the explicit channel reconstruction with reduced pilot and feedback overhead, and achieve better performance over state-of-the-art schemes in FDD massive MIMO-OFDM hybrid beamforming systems.}
\end{itemize}
\begin{itemize}
	\item To mitigate the performance loss caused by the phase quantization error brought by the practical low-resolution PSs, {\color{black} we introduce a transfer learning strategy  to adapt the E2E neural network to quantized PSs constraint.} Specifically, under the constraint of low-resolution PSs, we further fine-tune the E2E neural network based on a pre-trained network that assumes the ideal infinite-resolution PSs.
\end{itemize}
\begin{itemize}
	\item  In a hybrid MIMO system, uplink pilot combining and downlink pilot transmission need to meet the constant modulus constraint of the PSs. 	
	Therefore, we formulate the processing of uplink pilot combining or downlink pilot transmission as a single-layer fully-connected neural network, namely HPN. The learnable parameters are defined as the phase values of the PSs to meet the constant modulus constraint.
\end{itemize}

\textit{Notation}: This paper uses lower-case letters for scalars, lower-case bold face letters for column vectors, and upper-case bold face letters for matrices.
Superscripts $(\cdot)^*$, $(\cdot)^T$, $(\cdot)^H$, $(\cdot)^{-1}$, $(\cdot)^\dagger$ denote the conjugate, transpose, conjugate transpose, inversion, and Moore-Penrose inversion operators, respectively.
${\left\| {\mathbf{A}} \right\|_F}$ is the Frobenius norm of ${\mathbf{A}}$, respectively.
${{\rm{vec}}( {\mathbf{A}} )}$ and ${{\rm{angle}}( {\mathbf{A}} )}$ denote the vectorization operation and the phase values of ${\mathbf{A}}$, respectively.
${{\mathbf{I}}_n}$  denotes an identity matrix with size $n \times n$, while $\bm{1}_n$ ($\bm{0}_n$) denotes the vector of size $n$ with all the elements being $1$ ($0$).
${\Re\{\cdot\}}$ and ${\Im\{\cdot\}}$ denote the real part and imaginary part of the corresponding arguments, respectively.
$[\mathbf{A}]_{m,n}$ denotes
 the $m$th row and $n$th column element of $\mathbf{A}$, while $\mathbf{A}_{[:,m:n]}$
 is the sub-matrix containing the $m$th to $n$th columns of $\mathbf{A}$. The expectation is denoted by $\mathbb{E}(\cdot)$.

\section{System Model}\label{S2}

\begin{figure*}[!tp]
	\vspace{-10.0mm}
	\begin{center} 
		\includegraphics[width = 1\columnwidth, keepaspectratio]{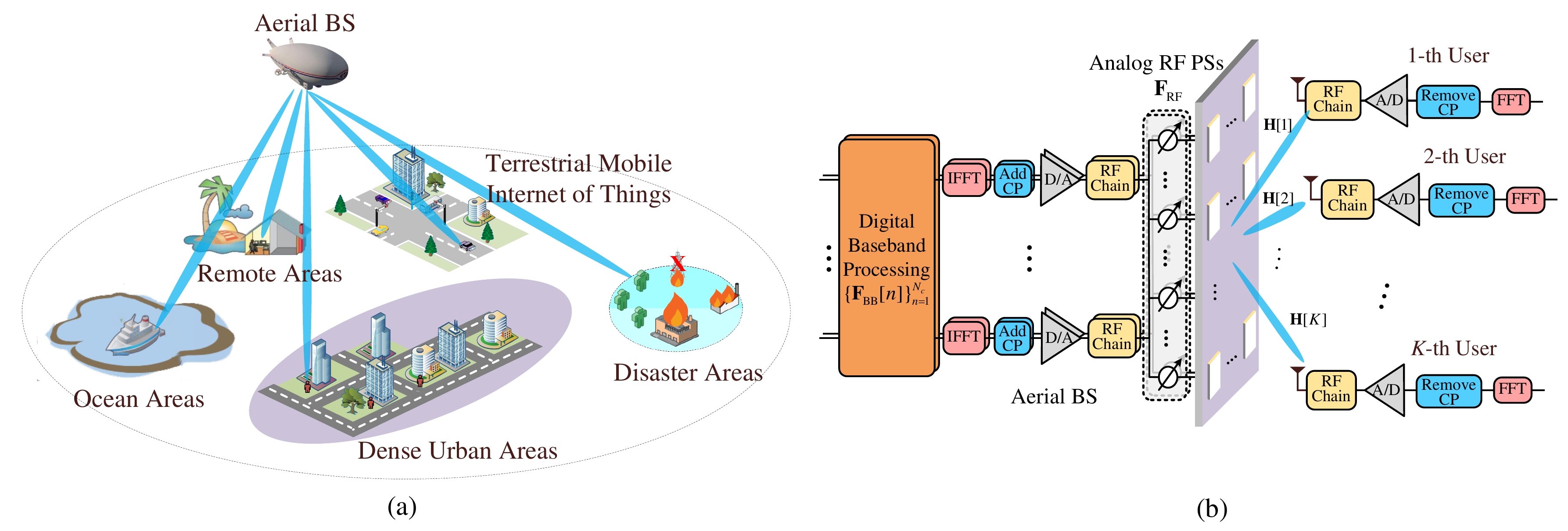}
	\end{center}
	\captionsetup{font = {footnotesize}, singlelinecheck = off, justification = raggedright, name = {Fig.}, labelsep = period}%
	\vspace{-6.0mm}
	\caption{(a) Aerial massive MIMO systems based on airships; (b) Block diagram of hybrid MIMO-OFDM system in the downlink.}
	\vspace{-6.0mm}
	\label{fig_A2G}
\end{figure*}

This paper investigates the multi-user hybrid beamforming for aerial massive MIMO-OFDM systems as shown in Fig.~\ref{fig_A2G}a, which can provide wide-area coverage for disaster areas, remote areas, ocean areas, dense urban areas, terrestrial mobile Internet of Things, etc. 
 Specifically, we consider the airship/balloon-based aerial BS adopts a uniform planar array (UPA) as shown in Fig.~\ref{fig_A2G}b, where the fully-connected hybrid MIMO architecture with $K$ radio frequency (RF) chains and $M$ antennas are employed. Moreover, we consider the aerial BS can simultaneously serve $K$ single-antenna terrestrial users, and the cyclic prefix (CP)-OFDM with $N_c$ subcarriers is used. In the downlink transmission, the transmit signal ${\mathbf x}[{n}]\in \mathbb{C}^{M\times 1}$ on the $n$-th subcarrier can be expressed as
\begin{equation}\label{form1}
	{\mathbf x}[{n}]=\sum_{k=1}^{K} {\mathbf F}_{\rm RF}{\mathbf f}_{\rm BB}[k,n]{ s}[{k,n}]={\mathbf F}_{\rm RF}{\mathbf F}_{{\rm BB}}[n]{\mathbf s}[{n}],
\end{equation}
where ${\mathbf f}_{\rm BB}[k,n]\in \mathbb{C}^{K\times 1}$ represents the digital baseband beamforming vector associated with the $k$-th terrestrial user on the $n$-th subcarrier, ${\mathbf F}_{{\rm BB}}[n]= \left[ {\mathbf f}_{\rm BB}[{1,n}],{\mathbf f}_{\rm BB}[{2,n}],\cdots,{\mathbf f}_{\rm BB}[{K,n}] \right] \in \mathbb{C}^{K\times K}$ is the total digital baseband beamforming matrix on the $n$-th subcarrier, ${\mathbf F}_{\rm RF}\in \mathbb{C}^{M\times K}$ is the RF analog beamforming matrix, ${ s}[{k,n}]$ is the data associated with the $k$-th terrestrial user delivered on the $n$-th subcarrier, and ${\mathbf s}[{n}] = \left[{ s}[{1,n}],{ s}[{2,n}],\cdots,{ s}[{K,n}]\right]^{T}\in \mathbb{C}^{K\times 1}$, where $\mathbb{E}\left({\mathbf s}[{n}]{\mathbf s}^H[{n}]\right)={{\mathbf{I}}_K}$ for $n=1,2,\cdots,N_{c}$. Moreover, since the analog beamforming matrix is implemented by the RF PSs, it  should meet the constant modulus constraint, i.e.,  $\left|\left[{\mathbf F}_{\rm RF} \right]_{i,j}\right|  =1,\forall i,j$. Finally, the product of the analog beamforming matrix and the digital beamforming matrix on each subcarrier should meet the power constraint, i.e., ${\left\| {\mathbf F}_{\rm RF}{\mathbf F}_{{\rm BB}}[n] \right\|_F^2 \leq P_t/N_c}$, and $P_t$ is the total transmit power. 

In the downlink data transmission stage, the signal ${y}[{k,n}]$ received at the $k$-th terrestrial user on the $n$-th subcarrier can be expressed as
\begin{align}\label{form2}
	{y}[{k,n}] = & {\mathbf h}^{H}[{k,n}]{\mathbf F}_{\rm RF}{\mathbf f}_{\rm BB}[{k,n}]{s}[{k,n}]  \\ \nonumber
	&+\sum_{k'\neq k}{\mathbf h}^{H}[{k,n}]{\mathbf F}_{\rm RF}{\mathbf f}_{\rm BB}[{k',n}]{s}[{k',n}]+{z}[{k,n}], \forall k,
\end{align}
where $k=1,2,\cdots K$, ${\mathbf h}[{k,n}]\in \mathbb{C}^{M\times 1}$ represents the downlink channel vector between the aerial BS and the $k$-th terrestrial user on the $n$-th subcarrier, and ${z}[{k,n}]\sim {\cal CN}\left( {0},\sigma_n^2 \right)$  is the additive white Gaussian noise (AWGN).

As for ${\mathbf h}[{k,n}], \forall k, n$, we consider the air-to-ground (A2G) channels between the aerial massive MIMO BS and the terrestrial users exhibit  a limited number of scatterers, where a typical multipath channel model \cite{ins9} is considered. Assuming that the number of downlink multipath components from the aerial BS to the $k$-th terrestrial user is $L_p$, the delay-domain channel can be expressed as
\begin{equation}\label{form3}
	\tilde{\mathbf h}(k,\tau)=\dfrac{1}{\sqrt{L_p}}\sum_{l=1}^{L_p}\alpha_{l,k}{\mathbf a}_t\left(\theta_{l,k},\phi_{l,k}\right)\delta\left(\tau-\tau_{l,k}\right),
\end{equation}
where $\alpha_{l,k}$ is the complex gain of the $l$-th path, $\tau_{l,k}$ is the delay of the $l$-th path, $\theta_{l,k}$ and $\phi_{l,k}$  are the azimuth and zenith angles of departures (AoDs) of the $l$-th path due to the UPA employed at the aerial BS, and ${\mathbf a}_t(\cdot)$ is the normalized transmit array response vector. Moreover, according to the time-frequency transformation, the $k$-th terrestrial user's frequency-domain channel on the $n$-th subcarier  can be expressed as
\begin{equation}\label{form4}
	{\mathbf h}[{k,n}]=\frac{1}{\sqrt{L_p}}\sum_{l=1}^{L_p}\alpha_{l,k}{\mathbf a}_t(\theta_{l,k},\phi_{l,k})e^{-{\rm j}\frac{2\pi n\tau_{l,k}}{N_cT_s}},
\end{equation}
where ${\rm j}=\sqrt{-1}$, and $T_s$ is the system's sampling interval.

 In the case of a UPA in the yz-plane, there are $N_{\rm y}$ and $N_{\rm z}$ antenna elements on the y and z axes, respectively, and $M=N_{\rm y}N_{\rm z}$ is the total number of antennas at the aerial BS. In this case, the array response vector can be expressed as
\begin{align}\label{form10}
	{\mathbf a}_t(\theta,\phi)=\left[ 1,\cdots,e^{{\rm j}\frac{2\pi}{\lambda}d(n\sin(\theta)\cos(\phi)+m\sin(\phi))}, 
	\cdots,e^{{\rm j}\frac{2\pi}{\lambda}d((N_{\rm y}-1)\sin(\theta)\cos(\phi)+(N_{\rm z}-1)\sin(\phi))} \right],
\end{align}
where $0\leq n< N_{\rm y}$ and $0\leq m< N_{\rm z}$ are respectively the y and z indices of the antenna elements, $\lambda$ and $d$ respectively denote the wavelength and adjacent antenna spacing, and $d = \frac{\lambda}{2}$.

\section{Proposed DL-Based Approach in TDD mode}\label{S_TDDDL}

\begin{figure*}[!tp]
	\vspace{-3.0mm}
	\begin{center} 
		\includegraphics[width = 1\columnwidth, keepaspectratio]{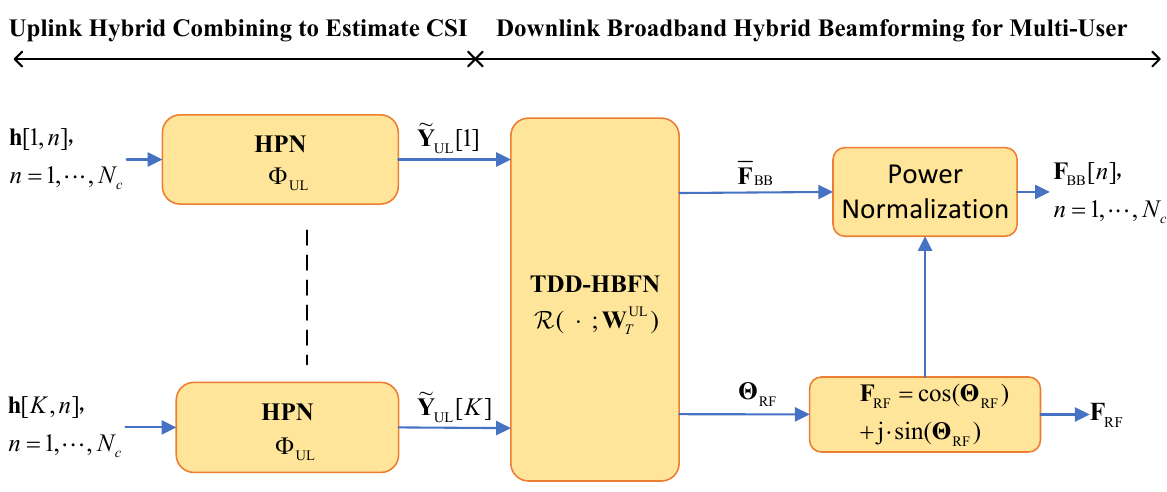}
	\end{center}
	\captionsetup{font = {footnotesize}, singlelinecheck = off, justification = raggedright, name = {Fig.}, labelsep = period}%
	\vspace{-1.5mm}
	\caption{Proposed DL-based joint pilot combining and multi-user hybrid beamforming design for TDD aerial massive MIMO-OFDM systems.}
	\vspace{-1.5mm}
	\label{fig_TDD}
\end{figure*}

In this section, we firstly present the processing procedure of the proposed DL-based joint design of hybrid combiner (i.e., pilot signals) in the uplink channel training stage and multi-user hybrid beamforming in the downlink data transmission stage for TDD mode. The proposed approach can be formulated as an E2E neural network as shown in Fig.~\ref{fig_TDD}, which consists of an HPN and a TDD-HBFN at the aerial BS, with the negative sum throughput as the loss function. In the following, we elaborate on how to model such a CSI acquisition and beamforming procedure as an E2E  neural network. 

\subsection{Processing Procedure and Problem Formulation in TDD Mode}\label{S5.TDD}

For TDD mode with hybrid analog-digital array at the aerial BS, estimating uplink high-dimensional CSI from a limited number of RF chains is a non-trivial problem and requires multiple pilot OFDM symbols.
Specifically, we consider that the terrestrial users adopt the mutually orthogonal pilot signals during the uplink pilot training stage, so the received pilot signals associated with different terrestrial users can be easily  distinguished at the aerial BS. For the $k$-th terrestrial user, the baseband pilot signal received at the aerial BS on the $n$-th subcarrier of the $q$-th pilot OFDM symbol can be expressed as 
\begin{equation}\label{form5_0}
	\tilde{\mathbf y}_{\rm UL}'[{q,k,n}]=\widetilde{\mathbf W}[q]{\mathbf h}[{k,n}]b[{k,q,n}]+\widetilde{\mathbf W}[q]\tilde{\mathbf z}_{\rm UL}'[{q,k,n}] \in \mathbb{C}^{K\times 1}, 1\leq q \leq Q,
\end{equation}
where $\tilde{\mathbf z}_{\rm UL}'[{q,k,n}]\sim {\cal CN}\left( \bm{0}_M,\sigma_n^2{{\mathbf{I}}_M} \right)$ is the AWGN, $\widetilde{\mathbf W}[q] \in \mathbb{C}^{K\times M}$ is the uplink combining matrix at the aerial BS, and $b[{k,q,n}]\in \mathbb{C}$ is the uplink pilot signal sent by the $k$-th terrestrial user.
Since the uplink combining matrix $\widetilde{\mathbf W}[q]$ is implemented through a fully connected PSs network, it should meet the constant modulus constraint, i.e., $\left| \left[ \tilde{\mathbf W}[q] \right]_{i,j}\right|=\sqrt{1/M}$. 
The aerial BS then further processes the received pilot signal as
\begin{equation}\label{form5_1}
	\tilde{\mathbf y}_{\rm UL}[{q,k,n}]=\tilde{\mathbf y}_{\rm UL}'[{q,k,n}]b^*[{k,q,n}]=\widetilde{\mathbf W}[q]{\mathbf h}[{k,n}]+\tilde{\mathbf z}_{\rm UL}[{q,k,n}] \in \mathbb{C}^{K\times 1},
\end{equation}
where $b[{k,q,n}]b^*[{k,q,n}] = 1$ \footnote{$\{b[{k,q,n}]\}_{n=1}^{N_c}$ here is a frequency-domain scrambling code in the pilot training stage, and the descrambling processing at the receiver can be implemented by multiplying the conjugate of $b[{k,q,n}]$ as shown in (\ref{form5_1}).  Due to $b[{k,q,n}]b^*[{k,q,n}] = 1$, the scrambling code $\{b[{k,q,n}]\}_{n=1}^{N_c}$ does not affect the subsequent signal processing, while introducing the scrambling code is necessary to ensure the low peak-to-average power ratio of the transmit signals in the time domain.},
 and $\tilde{\mathbf z}_{\rm UL}[{q,k,n}] = \tilde{\mathbf W}[q]\tilde{\mathbf z}_{\rm UL}'[{q,k,n}]b^*[{q,n}]\in \mathbb{C}^{K\times 1}$. By combining the baseband pilots received at the aerial BS in $Q$ pilot OFDM symbols, the received baseband pilot signal can be aggregately expressed as
\begin{equation}\label{form2_2}
	\tilde{\mathbf y}_{\rm UL}[{k,n}]=\widetilde{\mathbf W}{\mathbf h}[{k,n}]+\tilde{\mathbf z}_{\rm UL}[{k,n}],
\end{equation} 
where $\tilde{\mathbf y}_{\rm UL}[{k,n}] = \left[\tilde{\mathbf y}_{\rm UL}^T[{1,k,n}],\cdots,\tilde{\mathbf y}_{\rm UL}^T[{Q,k,n}]\right]^{T}\in \mathbb{C}^{QK \times 1}$, $\widetilde{\mathbf W} = \left[\widetilde{\mathbf W}^T[1],\cdots,\widetilde{\mathbf W}^T[Q]\right]^T\in \mathbb{C}^{QK \times M}$, and $\tilde{\mathbf z}_{\rm UL}[{k,n}] = \left[\tilde{\mathbf z}_{\rm UL}^T[{1,k,n}],\cdots,\tilde{\mathbf z}_{\rm UL}^T[{Q,k,n}]\right]^{T}\in \mathbb{C}^{QK \times 1}$. Finally, the aerial BS stacks the received pilot signals from all subcarriers, which can be expressed as
\begin{equation}\label{form2_3}
	\widetilde{\mathbf Y}_{\rm UL}[{k}]=\widetilde{\mathbf W}{\mathbf H}[{k}]+\widetilde{\mathbf Z}_{\rm UL}[{k}],
\end{equation}
where $\widetilde{\mathbf Y}_{\rm UL}[{k}] = \left[\tilde{\mathbf y}_{\rm UL}[{k,1}],\cdots,\tilde{\mathbf y}_{\rm UL}[{k,N_c}]\right]\in \mathbb{C}^{QK \times N_c}$, ${\mathbf H}[{k}] = \left[{\mathbf h}[{k,1}],\cdots,{\mathbf h}[{k,N_c}]\right]\in \mathbb{C}^{M \times N_c}$, and $\widetilde{\mathbf Z}_{\rm UL}[{k}] = [\tilde{\mathbf z}_{\rm UL}[{k,1}],\cdots,$ $\tilde{\mathbf z}_{\rm UL}[{k,N_c}]]\in \mathbb{C}^{QK \times N_c}$. Thanks to the reciprocity of the uplink channel and downlink channel in TDD mode, the aerial BS can obtain the downlink CSI based on the received uplink pilot signals for designing hybrid beamformer. This process can be expressed as 
\begin{equation}\label{form2_4}
	\left\{{\mathbf F}_{\rm RF},{\mathbf F}_{{\rm BB}}[1], \cdots, {\mathbf F}_{{\rm BB}}[N_c] \right\}={\cal R}\left(\widetilde{\mathbf Y}_{\rm UL}[{1}], \cdots, \widetilde{\mathbf Y}_{\rm UL}[{K}] \right),
\end{equation}
where ${\cal R}(\cdot)$ represents the mapping function from the received uplink pilot signals to the hybrid beamformer consisting of the analog part ${\mathbf F}_{\rm RF}$ and the digital part ${\mathbf F}_{{\rm BB}}[n]$ for $ 1\leq n\leq N_c$.   

According to the formula (\ref{form2}), the achievable rate of the $k$-th terrestrial user on the $n$-th subcarrier can be expressed as
\begin{equation}\label{form3}
	{ R}_{k,n} = {\rm log}_{2}\left( 1+\dfrac{\left| {\mathbf h}^H[{k,n}]{\mathbf F}_{\rm RF}{\mathbf f}_{\rm BB}[{k,n}] \right| ^{2}}{\sum_{k'\neq k}\left| {\mathbf h}^H[{k,n}]{\mathbf F}_{\rm RF}{\mathbf f}_{\rm BB}[{k',n}] \right| ^{2}+\sigma_n^2 } \right).
\end{equation}

Therefore, the sum rate $R$ in the downlink multi-user transmission can be expressed as
\begin{equation}\label{form_sumrate}
	{R} = \dfrac{1}{N_{c}}\sum_{k=1}^K\sum_{n=1}^{N_c}{R}_{k,n}.
\end{equation}

Based on the processing procedure aforementioned, the joint design of uplink pilot training and downlink multi-user hybrid beamforming in TDD mode can be formulated as the following optimization problem, i.e., 
\begin{align}
	\label{form9}
	\mathop{\arg\max}\limits_{\widetilde{\mathbf W},{\cal R(\cdot)}}\quad & R = \dfrac{1}{N_{c}} \sum\limits_{k=1}^K\sum\limits_{n=1}^{N_c}  
	{\rm log}_{2}\left( 1+\dfrac{\big| {\mathbf h}^H[{k,n}]{\mathbf F}_{\rm RF}{\mathbf f}_{\rm BB}[{k,n}] \big| ^{2}}{\sum_{k'\neq k}\big| {\mathbf h}^{H}[{k,n}]{\mathbf F}_{\rm RF}{\mathbf f}_{\rm BB}[{k',n}] \big| ^{2}+\sigma_n^2 } \right), \nonumber \\
	{\rm s.t.} \quad &  \left\{ {\mathbf F}_{\rm RF},{\mathbf F}_{{\rm BB}}[1], \cdots, {\mathbf F}_{{\rm BB}}[N_c] \right\}  = {\cal R}\left(\widetilde{\mathbf Y}_{\rm UL}[{1}], \cdots, \widetilde{\mathbf Y}_{\rm UL}[{K}]\right), \nonumber \\
	& \left| \left[ {\mathbf F}_{\rm RF} \right]_{i,j}\right|  =1,\forall i,j,  \nonumber \\
	& \left\| {\mathbf F}_{\rm RF}{\mathbf F}_{{\rm BB}}[n] \right\|_{F}^{2}  \leq P_t/N_c,\forall n,  \nonumber \\
	& \widetilde{\mathbf Y}_{\rm UL}[{k}]=\widetilde{\mathbf W}{\mathbf H}[{k}]+\widetilde{\mathbf Z}_{\rm UL}[{k}],\forall k, \nonumber \\
	& \left| \left[ \widetilde{\mathbf W} \right]_{i,j}\right|  =\sqrt{1/M}, \forall i,j.
\end{align}

 By contrast, the conventional approaches for TDD systems typically decompose the downlink multi-user transmission into two separated and mutually independent modules: i.e., uplink channel estimation and downlink hybrid beamforming, which would inevitably introduce performance loss. Jointly designing the channel estimation and hybrid beamforming as shown in formula (\ref{form9}) is expected to achieve better performance, but how to solve such an optimization problem with complicated constraints and a large number of variables is challenging.
To this end, by modeling the uplink channel estimation and downlink multi-user hybrid broadband beamforming as an E2E neural network, we propose a DL-based approach to achieve the joint optimization.


\subsection{Uplink Pilot Combining for Hybrid MIMO-OFDM Architecture}\label{S5.TDD_PT}

In the uplink channel training stage, we consider the aerial BS receives $Q$ pilot OFDM symbols and combines them to obtain the received baseband pilot signals, i.e., the measurements can be expressed as $\tilde{\mathbf y}_{\rm UL}[{k,n}]=\widetilde{\mathbf W}{\mathbf h}[{k,n}]+\tilde{\mathbf z}_{\rm UL}[{k,n}],$ for $1\leq k \leq K,1\leq n \leq N_c$. 

\begin{figure*}[!tp]
	\vspace{-4.0mm}
	\captionsetup{font={footnotesize}, name = {Fig.}, labelsep = period} 
	\captionsetup[subfigure]{singlelinecheck = on, justification = raggedright, font={footnotesize}}
	\centering
	
	\subfloat[]{
		\label{TDD_HPN}
		\begin{minipage}[t]{0.45\linewidth}
			\centering
			\includegraphics[scale=0.94]{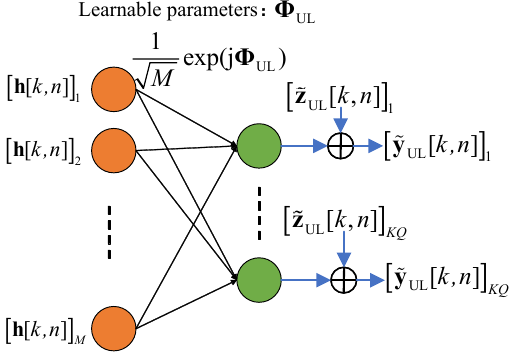}
		\end{minipage}
	}
	\subfloat[]{
		\label{FDD_HPN}
		\begin{minipage}[t]{0.55\linewidth}
			\centering
			\includegraphics[scale=0.94]{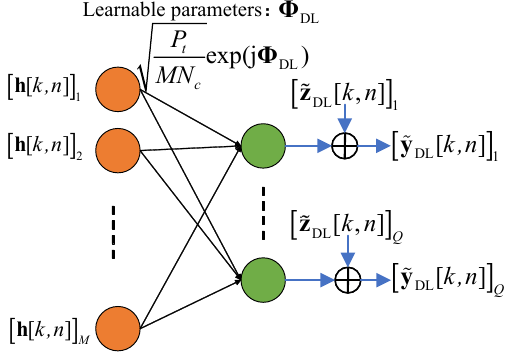}
		\end{minipage}
		
	}
	\vspace{-0.0mm}
	\caption{Proposed HPN for (a) uplink hybrid combining at the aerial BS in TDD systems, and (b) downlink hybrid pilot transmission at the aerial BS in FDD systems.}
	\vspace{-3.0mm}
	\label{fig_HPN}
\end{figure*}

As for the narrowband fully-digital MIMO systems, this process can be modeled as a fully-connected dimensionality-reduction linear network, where the combining matrix is the weight matrix of the dense layer. By contrast, as for the hybrid MIMO deployed at the aerial BS in our work, we consider all the RF chains are activated for the uplink pilot combining, and the combining matrix should meet the constant modulus constraint, i.e., $\big| \left[ \widetilde{\mathbf W} \right]_{i,j} \big| ={\frac{1}{\sqrt{M}}}$, $\forall i$, $j$, leading the conventional linear fully-connected layer in \cite{Yuwei} to be ineffective. To this end, we propose an HPN as shown in Fig.~\ref{TDD_HPN}, where the phase matrix ${\mathbf \Phi}_{\rm UL} = {\rm angle}({\widetilde{\mathbf W}})$ can be learned and determined in the deep learning training stage, and each element of ${\mathbf \Phi}_{\rm UL}$ represents the phase value of the corresponding element in $\widetilde{\mathbf W}$. In this way, we have 
\begin{equation}\label{form12}
	\widetilde{\mathbf W}=\frac{1}{\sqrt{M}}{\rm exp}\left({{\rm j}{\mathbf \Phi}_{\rm UL}}\right)=\frac{1}{\sqrt{M}} \left(\cos({\mathbf \Phi}_{\rm UL})+{\rm j} \sin({\mathbf \Phi}_{\rm UL})\right).
\end{equation}


{\color{black} Note that, for most of the existing deep learning frameworks, it is difficult to directly impose the complex constant modulus constraint on the neural network weight parameters (i.e., the combining matrix $\widetilde{\mathbf W}$), thus we choose the phase matrix ${\mathbf \Phi}_{\rm UL}$ instead of $\widetilde{\mathbf W}$ as the trainable parameters of the HPN and impose the complex exponential function ${\rm exp(j \cdot)}$ on it to indirectly meet the complex constant modulus constraint in hybrid MIMO architecture. }

\subsection{Downlink Broadband Hybrid Beamforming for Multi-User}\label{S5.TDD_HP}

\begin{figure*}[t]
	\vspace{-4.0mm}
	\begin{center} 
		\includegraphics[width = 1\columnwidth, keepaspectratio]{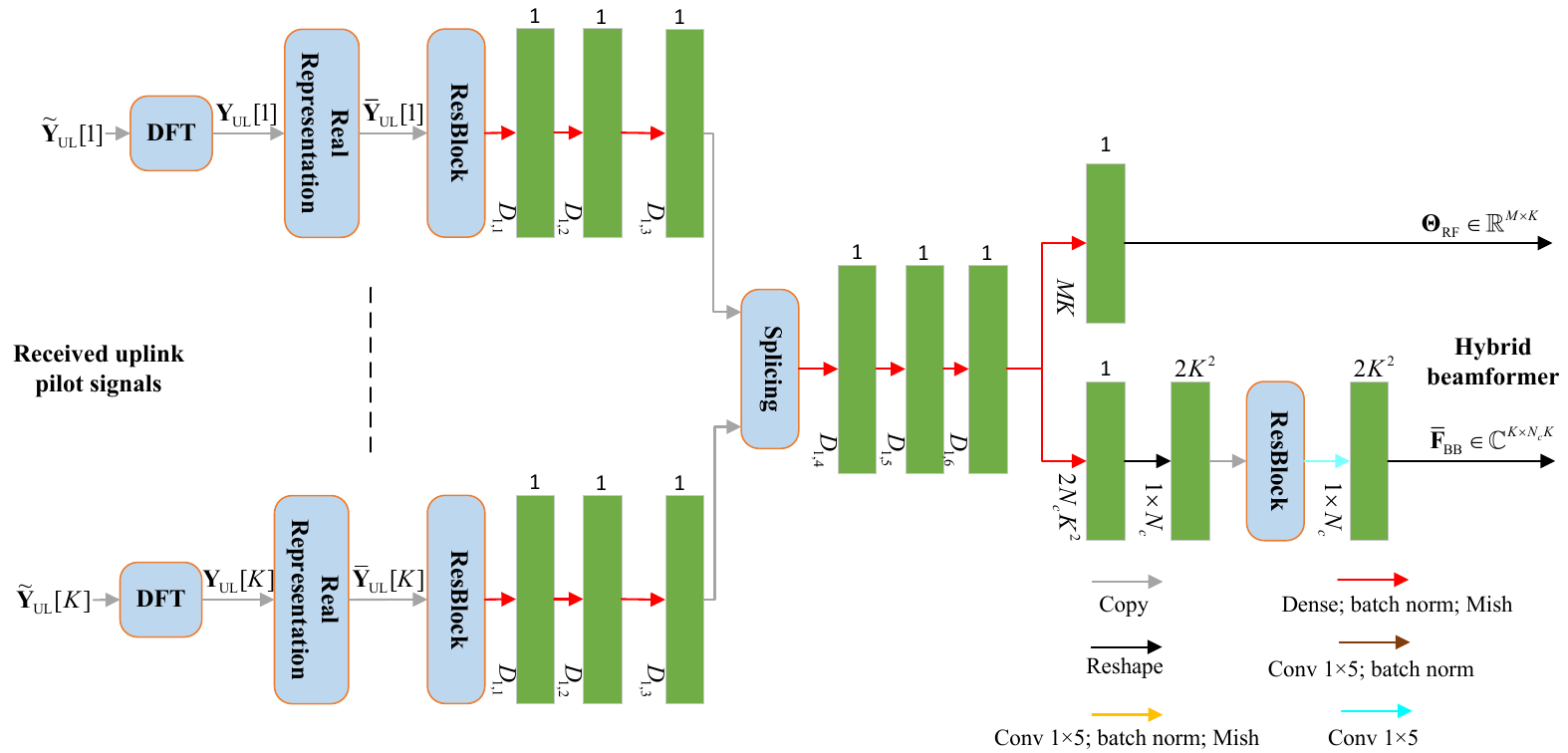}
	\end{center}
	\captionsetup{font = {footnotesize}, singlelinecheck = off, justification = raggedright, name = {Fig.}, labelsep = period}%
	\vspace{-6.0mm}
	\caption{Proposed TDD-HBFN for downlink multi-user broadband hybrid beamforming at the aerial BS.}
	\vspace{-6.0mm}
	\label{fig_TDDHB}
\end{figure*}

In TDD mode, we consider that the aerial BS can perform hybrid beamforming according to the implicit CSI extracted from the received baseband uplink pilot signals. We model this process as a TDD-HBFN as shown in Fig.~\ref{fig_TDDHB}. Owing to the typical A2G channels exhibiting sparsity in both the delay domain and angle domain, $\tilde{\mathbf Y}_{\rm UL}[{k}]$ can be further compressed in the delay domain. Therefore, we perform DFT on $\tilde{\mathbf Y}_{\rm UL}[{k}]$ to transform it from the frequency domain to the delay domain, i.e., 
\begin{equation}\label{form13}
	{\mathbf Y}_{\rm UL}[k] = {\mathbf F}_{N_c}\widetilde{\mathbf Y}^H_{\rm UL}[k],
\end{equation}
where ${\mathbf F}_{N_c}\in \mathbb{C}^{N_c\times N_c}$ is the DFT matrix of $N_c$ points. Such a compressibility of the received pilot signal ${\mathbf Y}_{\rm UL}[k]$ in the delay domain indicates the local spatial correlation among adjacent neurons (such as convolutional layers in the neural network). Since neural networks are more effective for real-valued operations than complex-valued operations, the TDD-HBFN reshapes ${\mathbf Y}_{\rm UL}[k]$ into a real-valued tensor $\bar{\mathbf Y}_{\rm UL}[k]\in \mathbb{R}^{2KQ\times 1 \times N_c}$, which can be expressed as
\begin{equation}\label{form14}
	\begin{cases}
	\left[ \bar{\mathbf Y}_{\rm UL}[k]\right]_{[1:KQ,1,1:N_c]}=\Re\{{{\mathbf Y}_{\rm UL}^{T}[k]}\},\\
	\left[ \bar{\mathbf Y}_{\rm UL}[k]\right]_{[KQ+1:2KQ,1,1:N_c]}=\Im\{{{\mathbf Y}_{\rm UL}^{T}[k]}\}.
\end{cases}
\end{equation}

\begin{figure*}[t]
 	\vspace{-9.0mm}
 	\begin{center} 
 		\includegraphics[width = 0.7\columnwidth, keepaspectratio]{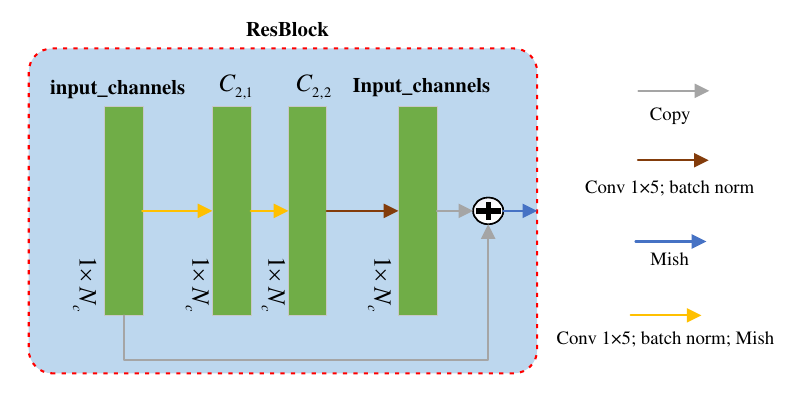}
 	\end{center}
 	\captionsetup{font = {footnotesize}, singlelinecheck = off, justification = raggedright, name = {Fig.}, labelsep = period}%
 	\vspace{-7mm}
 	\caption{Illustration of the proposed ResBlock.}
 	\vspace{-7.0mm}
 	\label{fig_resblock}
 \end{figure*}

To extract the local correlation of the received pilot signal ${\bar{\mathbf Y}_{\rm UL}[k]}$ in the delay domain, the convolution kernel size of each convolutional layer is set to $1 \times 5$. As shown in Fig.~\ref{fig_TDDHB}, $\bar{\mathbf Y}_{\rm UL}[k]$  is first input to a ``ResBlock unit'' as shown in Fig.~\ref{fig_resblock} for primary feature extraction. 
Each ResBlock has four layers: the first layer is the input layer; the second and third layers are convolutional layers, which generate $C_1$ and $C_2$ feature maps, respectively; and the last layer is also a convolutional layer, which generates the same number of feature maps as the first layer. 
 Each convolutional layer adopts appropriate zero padding so that the dimension $1\times N_c$ of the input feature maps will not be changed by convolutional layers. Besides, we introduce a residual connection to transfer the input layer data stream directly to the fourth layer in each ResBlock. This method is developed from the deep residual network. It can avoid the vanishing gradient problem caused by multiple stacked non-linear transformations when the number of neural network layers is large  \cite{res}. Then, we adopt three hidden dense layers whose numbers of neurons are $D_{1,1}$, $D_{1,2}$, and $D_{1,3}$, respectively, to extract each terrestrial user's implicit CSI vector with dimension $D_{1,3}\times 1$, and then splice these vectors into an aggregated vector containing all $K$ users' implicit CSI  with dimension $KD_{1,3}\times 1$. The TDD-HBFN processes the spliced $K$ users' implicit CSI vector through four dense layers with the numbers of neurons $D_{1,4}$, $D_{1,5}$, $D_{1,6}$, and $2N_cK^2+MK$, respectively. The output of the fourth dense layer is divided into two parts. The first part aims to obtain the phase values ${\mathbf \Theta}_{\rm RF}\in \mathbb{R}^{M\times K}$ of the analog beamformer, and the remaining part is input to a ResBlock to obtain the digital beamformer. Finally, the output of the TDD-HBFN can be expressed as
\begin{align}\label{form17}
	\{ {\mathbf \Theta}_{\rm RF}, {\mathbf {\overline F}}_{{\rm BB}}\}
 ={\cal R}\left(\widetilde{\mathbf Y}_{\rm UL}[1],\cdots,\widetilde{\mathbf Y}_{\rm UL}[K];{\mathbf W}_{ T}^{\rm UL}\right),
\end{align}
where ${\mathbf {\overline F}}_{{\rm BB}} = \left[{\mathbf {\overline F}}_{{\rm BB}}[1],\cdots,{\mathbf {\overline F}}_{{\rm BB}}[N_c]\right]\in \mathbb{C}^{K\times N_cK}$, ${\mathbf {\overline F}}_{{\rm BB}}[n]\in \mathbb{C}^{K\times K}$ is the digital beamformer without power normalization on the $n$-th subcarrier, ${\cal R}(\cdot;{\mathbf W}_{ T}^{\rm UL})$ is the mapping function from the input to the output of the TDD-HBFN, and ${\mathbf W}_{ T}^{\rm UL}$ is the set of the learnable parameters of the TDD-HBFN. 

We use the Mish activation function \cite{Mish} in each hidden layer to provide nonlinear mapping ability for the network. 
To speed up the convergence of the network and avoid overfitting, we introduce the batch normalization between each hidden layer and activation function. In addition, we obtain the phase values of the analog beamforming matrix through the TDD-HBFN instead of the real and imaginary parts to meet the constant modulus constraint. In this way, the analog beamformer obtained by the TDD-HBFN can be expressed as
\begin{align} \label{form17}
{\mathbf F}_{\rm RF} & = \cos({\mathbf \Theta}_{\rm RF})+{\rm j}\cdot \sin({\mathbf \Theta}_{\rm RF}).
\end{align}

Finally, we need to impose power constraint on the digital beamformer, i.e.,
\begin{align}\label{form18}
	{\mathbf F}_{{\rm BB}}[n]=\min(\sqrt{P_t/N_c},\left\| {\mathbf F}_{\rm RF}{\mathbf{ \overline{F}}}_{{\rm BB}}[n]\right\|_{F})\frac{{\mathbf{ \overline{F}}}_{{\rm BB}}[n]}{\left\| {\mathbf F}_{\rm RF}{\mathbf{ \overline{F}}}_{{\rm BB}}[n]\right\|_{F}}, 1\leq n \leq N_c.
\end{align}

The characteristics of sparse connectivity and weight sharing of the convolutional layers can efficiently extract the spatial correlation of the processed data, which will improve the generalization ability \cite{CNN} of neural networks. Therefore, we adopt convolutional layers to take advantage of the received pilot signals' correlation in both the frequency domain and delay domain, so that the generalization ability of the proposed neural network can be enhanced to alleviate overfitting.

In sum, the learnable parameters of the proposed E2E model in TDD mode include the phase matrix ${\mathbf \Phi}_{\rm UL}$ of the HPN and the learnable parameter set ${\mathbf W}_{ T}^{\rm UL}$ of the TDD-HBFN.

\section{Proposed DL-Based Approach in FDD mode}\label{S5}
\begin{figure*}[!tp]
	\vspace{-10.0mm}
	\begin{center} 
		\includegraphics[width = 1\columnwidth, keepaspectratio]{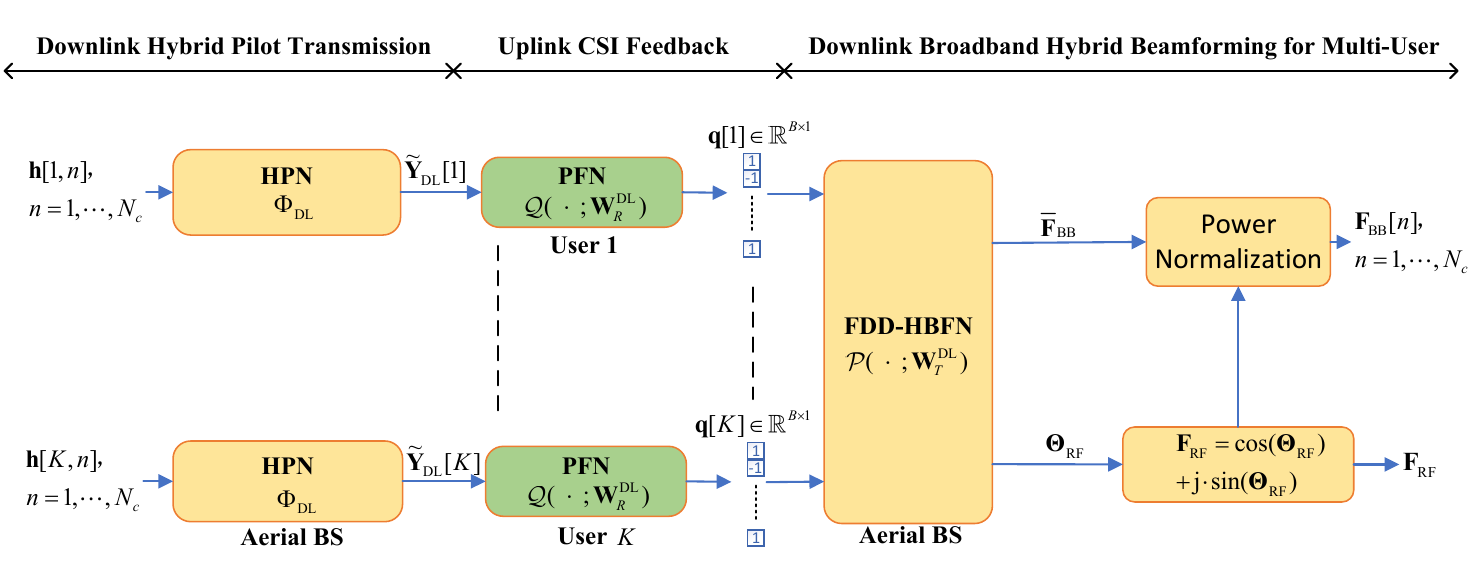}
	\end{center}
	\captionsetup{font = {footnotesize}, singlelinecheck = off, justification = raggedright, name = {Fig.}, labelsep = period}%
	\vspace{-8.0mm}
	\caption{Proposed DL-based joint downlink pilot signals, uplink CSI feedback, and multi-user hybrid beamforming design for FDD aerial massive MIMO-OFDM systems, where the yellow and green block diagrams represent the modules processed at the aerial BS and the terrestrial users, respectively.}
	\vspace{-7.0mm}
	\label{FIG_sys2}
\end{figure*}

In this section, we extend the proposed DL-based E2E approach from TDD mode to FDD mode, as shown in Fig.~\ref{FIG_sys2}, which consists of an HPN at the aerial BS, a PFN at the terrestrial users, and an FDD-HBFN at the aerial BS. Note that due to the non-reciprocity of the uplink/downlink channels, the estimated downlink CSI or the received downlink pilot signals require to be fed back to the aerial BS for achieving the downlink CSI at the aerial BS. Nevertheless, as the capacity of feedback link is limited and the channel coherence time is insufficient, the CSI feedback has to trade the feedback accuracy with feedback latency, since the prohibitively high CSI feedback time overhead would lead to the outdated CSI at the aerial BS and the deteriorated downlink payload spectral efficiency.

\subsection{Processing Procedure and Problem Formulation in FDD Mode}\label{S5.0}
 In the downlink pilot training stage, we assume that the aerial BS transmits $Q$ pilot OFDM symbols. The pilot signal transmitted by the aerial BS can be denoted as $\tilde{\mathbf x}[q]g[{q,n}]\in \mathbb{C}^{M\times 1}$ for $1 \leq q \leq Q$, where $\tilde{\mathbf x}[q]\in \mathbb{C}^{M\times 1}$ is the RF pilot signal and $g[{q,n}]\in \mathbb{C}$ is the baseband pilot signal. The $k$-th terrestrial user's received pilot signal, denoted by $\tilde{y}_{\rm DL}'[{q,k,n}]\in \mathbb{C}$, can be expressed as
\begin{equation}\label{form6}
	\tilde{y}_{\rm DL}'[{q,k,n}]={\mathbf h}^{H}[{k,n}]\tilde{\mathbf x}[q]g[{q,n}]+\tilde{ z}_{\rm DL}'[{q,k,n}],
\end{equation}
where $\tilde{ z}_{\rm DL}'[{q,k,n}]\sim {\cal CN}\left( {0},\sigma_n^2\right)$ is the AWGN. After receiving the pilot signal $\tilde{y}_{\rm DL}'[{q,k,n}]$, the $k$-th user performs further processing, which yields
\begin{equation}\label{form4_7}
	\tilde{y}_{\rm DL}[{q,k,n}]=\tilde{y}_{\rm DL}'[{q,k,n}]g^*[{q,n}]={\mathbf h}^{H}[{k,n}]\tilde{\mathbf x}[q]+\tilde{ z}_{\rm DL}[{q,k,n}],
\end{equation}
where $g[{q,n}]g^*[{q,n}] = 1$ 
\footnote{$\{g[{q,n}]\}_{n=1}^{N_c}$ here is also a frequency-domain scrambling code in the pilot training stage, which is used to reduce the peak-to-average power ratio of the transmitted signal in the time domain.}
 and $\tilde{ z}_{\rm DL}[{q,k,n}] = \tilde{ z}_{\rm DL}'[{q,k,n}]g[{q,n}]^*$. Note that since we adopt a fully-connected hybrid MIMO architecture, the pilot signal transmitted by the aerial BS can be divided into the RF pilot signal and the baseband pilot signal, where the RF pilot signal should meet the constant modulus constraint, i.e., $\left| \left[ \tilde{\mathbf x} \right]_{i}\right|=\sqrt{\frac{P_t}{MN_c}} $. By combining the received signals of $Q$ pilot OFDM symbols together, the aggregated pilot signal received at the $k$-th terrestrial user on the $n$-th subcarrier can be expressed as 
\begin{equation}\label{form4_8}
	\tilde{\mathbf y}_{\rm DL}[{k,n}]=\widetilde{\mathbf X}{\mathbf h}[{k,n}]+\tilde{\mathbf z}_{\rm DL}[{k,n}],
\end{equation}
where $\tilde{\mathbf y}_{\rm DL}[{k,n}] = \left[\tilde{y}_{\rm DL}[{1,k,n}],\cdots,\tilde{y}_{\rm DL}[{Q,k,n}]\right]^H\in \mathbb{C}^{Q \times 1}$, 
$\widetilde{\mathbf X} =  \left[\tilde{\mathbf x}[1], \\ \cdots,\tilde{\mathbf x}[Q]\right]^{H}\in \mathbb{C}^{Q \times M}$, and $\tilde{\mathbf z}_{\rm DL}[{k,n}] =  \left[\tilde{ z}_{\rm DL}[{1,k,n}],\cdots,\tilde{ z}_{\rm DL}[{Q,k,n}]\right]^{H}\in \mathbb{C}^{Q \times 1}$. By collecting the received pilot signals $\tilde{\mathbf y}_{\rm DL}[{k,n}]$ for $n=1,2,\cdots,$ $N_{c}$, we can finally obtain the received pilot signals over all $N_c$ subcarriers, i.e., $\widetilde{\mathbf Y}_{\rm DL}[{k}] = [\tilde{\mathbf y}_{\rm DL}[{k,1}],$ $\cdots,\tilde{\mathbf y}_{\rm DL}[{k,N_{c}}]]\in \mathbb{C}^{Q \times N_c}$, which can be expressed as

\begin{equation}\label{form8}
	\widetilde{\mathbf Y}_{\rm DL}[{k}]=\widetilde{\mathbf X}{\mathbf H}[{k}]+\widetilde{\mathbf Z}_{\rm DL}[{k}],
\end{equation}
where $\widetilde{\mathbf Z}_{\rm DL}[{k}] = [\tilde{\mathbf z}_{\rm DL}[{k,1}],\cdots,\tilde{\mathbf z}_{\rm DL}[{k,N_c}]]\in \mathbb{C}^{Q \times N_c}$.

In the feedback stage, we consider the $k$-th terrestrial user extracts the implicit CSI from $\widetilde{\mathbf Y}_{\rm DL}[{k}]$ and compresses it 
into $B$ bits, which are then fed back to the aerial BS. This process can be mathematically expressed as  
\begin{equation}\label{form6}
	{\mathbf q}[{k}]={\cal Q}(\widetilde{\mathbf Y}_{\rm DL}[{k}])\in \mathbb{R}^{B \times 1},
\end{equation} 
where ${\mathbf q}[{k}]$ is the $k$-th terrestrial user's binary bit vector fed back to the aerial BS, and the quantizer ${\cal Q}(\cdot)$ is a mapping function from the received pilot signal $\widetilde{\mathbf Y}_{\rm DL}[{k}]$ to the feedback bit vector ${\mathbf q}[{k}]$.

As for the $K$ terrestrial users scheduled in the same group for multi-user broadband hybrid beamforming, the aerial BS will collect the $KB$ feedback bits from the $K$ terrestrial users to obtain the aggregated binary bit vector, denoted by  ${\mathbf q}=\left[{\mathbf q}[{1}]^T,...,{\mathbf q}[{K}]^T\right]^T\in \mathbb{R}^{KB \times 1}$. Based on ${\mathbf q}$, the aerial BS designs a multi-user hybrid broadband beamformer, and this process can be expressed as
\begin{equation}\label{form7}
	\left\{{\mathbf F}_{\rm RF},{\mathbf F}_{{\rm BB}}[1], {\mathbf F}_{{\rm BB}}[2], \cdots, {\mathbf F}_{{\rm BB}}[N_c] \right\}={\cal P}(\mathbf q),
\end{equation} 
where the function ${\cal P}(\cdot)$ represents the mapping function from the $K$ terrestrial users' feedback bit vector ${\mathbf q}$ to the hybrid beamforming matrix, including the analog part ${\mathbf F}_{\rm RF}$ and the digital part ${\mathbf F}_{{\rm BB}}[n]$ for $ 1\leq n \leq N_c$.  

Based on the above processing procedure and the optimization objective of formula (\ref{form_sumrate}), the joint design of downlink pilot signals, uplink CSI feedback, and broadband multi-user hybrid beamforming can be formulated as
	\begin{align}
		\label{form88}
		\mathop{\arg\max}\limits_{\widetilde{\mathbf X},{\cal Q(\cdot)},{\cal P(\cdot)}}\quad & R = \dfrac{1}{N_{c}} \sum\limits_{k=1}^K\sum\limits_{n=1}^{N_c}  
		{\rm log}_{2}\left( 1+\dfrac{\big| {\mathbf h}^H[{k,n}]{\mathbf F}_{\rm RF}{\mathbf f}_{\rm BB}[{k,n}] \big| ^{2}}{\sum_{k'\neq k}\big| {\mathbf h}^{H}[{k,n}]{\mathbf F}_{\rm RF}{\mathbf f}_{\rm BB}[{k',n}] \big| ^{2}+\sigma_n^2 } \right), \nonumber \\
		{\rm s.t.} \quad &  \left\{ {\mathbf F}_{\rm RF},{\mathbf F}_{{\rm BB}}[1], \cdots, {\mathbf F}_{{\rm BB}}[N_c]\right\}  = {\cal P}\left({\mathbf q} \right), \nonumber \\
		& {\mathbf q}[{k}]  ={\cal Q}\left(\widetilde{\mathbf Y}_{\rm DL}[{k}]\right),\forall k,  \nonumber \\
		& \left| \left[ {\mathbf F}_{\rm RF} \right]_{i,j}\right|  =1,\forall i,j,  \nonumber \\
		& \left\| {\mathbf F}_{\rm RF}{\mathbf F}_{{\rm BB}}[n] \right\|_{F}^{2}  \leq P_t/N_c,\forall n,  \nonumber \\
		& \widetilde{\mathbf Y}_{\rm DL}[{k}]=\widetilde{\mathbf X}{\mathbf H}[{k}]+\widetilde{\mathbf Z}_{\rm DL}[{k}],\forall k, \nonumber \\
		& \left| \left[ \widetilde{\mathbf X} \right]_{i,j}\right|  =\sqrt{\dfrac{P_t}{MN_c}}, \forall i,j.
	\end{align}
	
Formula (\ref{form88}) is a complicated joint optimization problem, and how to solve this problem is challenging. Therefore, we propose a DL-based E2E approach to jointly optimize the downlink pilot signals, uplink CSI feedback, and multi-user hybrid broadband beamforming for FDD aerial massive MIMO-OFDM systems.

\subsection{Downlink Pilot Design for Hybrid MIMO-OFDM Architecture }\label{S4.1}
In the downlink channel training stage, we consider the aerial BS sends the training signals in $Q$ pilot OFDM symbols, i.e., $\widetilde{\mathbf X}$, and the received pilot signals are $\tilde{\mathbf y}_{\rm DL}[{k,n}]=\widetilde{\mathbf X}{\mathbf h}[{k,n}]+\tilde{\mathbf z}_{\rm DL}[{k,n}],$ for $1\leq k \leq K,1\leq n \leq N_c$ as obtained in (\ref{form4_8}). 
This process can be modeled by an HPN as shown in Fig.~\ref{FDD_HPN}. In FDD systems, the output of the HPN is the downlink pilot signals $\tilde{\mathbf y}_{\rm DL}[{k,n}]$ received at the terrestrial user and the trainable parameters can be regarded as the phase values ${\mathbf \Phi}_{\rm DL}$ of the downlink training signals.
The training signals sent by the aerial BS can be expressed as
\begin{equation}\label{form12}
	\widetilde{\mathbf X}=\sqrt{\dfrac{P_t}{MN_c}}{\rm exp}\left({{\rm j}{\mathbf \Phi}_{\rm DL}}\right)=\sqrt{\dfrac{P_t}{MN_c}}(\cos({\mathbf \Phi}_{\rm DL})+{\rm j} \sin({\mathbf \Phi}_{\rm DL})).
\end{equation}

\begin{figure*}[!tp]
	\vspace{-3.0mm}
	\begin{center} 
		\includegraphics[width = 0.8\columnwidth, keepaspectratio]{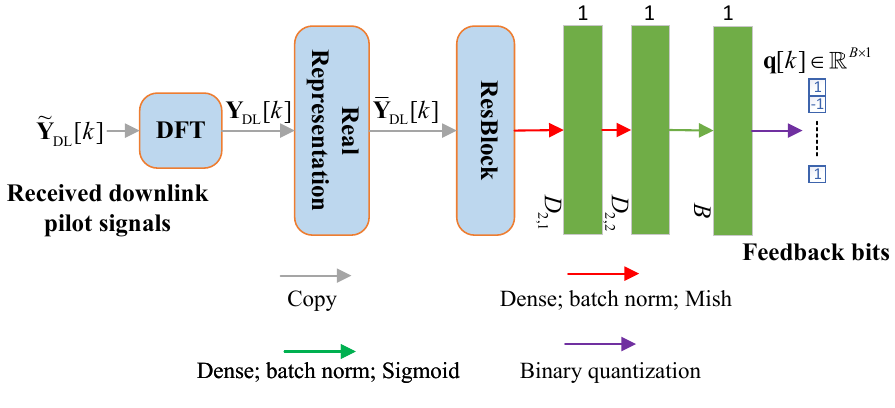}
	\end{center}
	\captionsetup{font = {footnotesize}, singlelinecheck = off, justification = raggedright, name = {Fig.}, labelsep = period}%
	\vspace{-5.0mm}
	\caption{Proposed PFN for uplink CSI feedback at the terrestrial users in FDD aerial massive MIMO-OFDM systems.}
	\vspace{-6.0mm}
	\label{fig_pilot_feedback}
\end{figure*}

\subsection{Uplink CSI Feedback}\label{S4.2}

We consider the uplink CSI feedback process of the $k$-th terrestrial user. The $k$-th terrestrial user needs to extract the implicit CSI from the received pilot signal $\widetilde{\mathbf Y}_{\rm DL}[{k}]$ and compress it into $B$ bits, which are then fed back to the aerial BS. We model this process as a PFN, as shown in Fig.~\ref{fig_pilot_feedback}. The PFN first transforms the received pilot signals from the frequency domain to the delay domain via DFT, i.e., 
\begin{equation}\label{form13}
	{\mathbf Y}_{\rm DL}[{k}] = {\mathbf F}_{N_c}\widetilde{\mathbf Y}_{\rm DL}[{k}],
\end{equation}
where ${\mathbf Y}_{\rm DL}[{k}]\in \mathbb{C}^{N_c\times Q}$ is the received pilot signal in the delay domain. Then, the PFN reshapes the delay domain pilot signal ${\mathbf Y}_{\rm DL}[{k}]$ into a real-valued tensor $\bar{\mathbf Y}_{\rm DL}[k]\in \mathbb{R}^{2Q\times 1 \times N_c}$
\begin{equation}\label{form4_3}
	\begin{cases}
		\left[ \bar{\mathbf Y}_{\rm DL}[k]\right]_{[1:Q,1,1:N_c]}=\Re\{{{\mathbf Y}_{\rm DL}^{T}[k]}\},\\
		\left[ \bar{\mathbf Y}_{\rm DL}[k]\right]_{[Q+1:2Q,1,1:N_c]}=\Im\{{{\mathbf Y}_{\rm DL}^{T}[k]}\}.
	\end{cases}
\end{equation}

Similar to the TDD-HBFN proposed in Section \ref{S_TDDDL}, the PFN mainly uses one-dimensional convolution to extract the local sparse features of the received pilot signal  $\bar{\mathbf Y}_{\rm DL}[k]$ represented in the delay domain. As shown Fig.~\ref{fig_pilot_feedback}, the PFN firstly processes $\bar{\mathbf Y}_{\rm DL}[k]$ by using a ResBlock and the following two dense layers (with $D_{2,1}$ and $D_{2,2}$ neurons, respectively) for implicit CSI extraction, and then the PFN adopts an output dense layer for compressing the implicit CSI to a binary bit vector due to the limited feedback link capacity. In the output layer, we use the sigmoid function and subtract 0.5 to limit the output to the range of $[-0.5,0.5]$, then convert them to $B$ feedback bits through a sign function $\rm sign(\cdot)$.  As a discontinuous function, $\rm sign(\cdot)$ is non-derivative, so we set the gradient of the quantization function to 1 for facilitating the backpropagation of the gradient in the training phase \cite{quancsinet}. The $B$ feedback bits can be expressed as
\begin{align}\label{form4_4}
	{\mathbf q}[k] ={\cal Q}(\widetilde{\mathbf Y}_{\rm DL}[k];{\mathbf W}_{R}^{\rm DL}),
\end{align}
where ${\mathbf W}_{R}^{\rm DL}$ is the set of the learnable parameters of the PFN, ${\mathbf q}[k]$ is the feedback bit vector of the $k$-th terrestrial user, and ${\cal Q}(\cdot;{\mathbf W}_{R}^{\rm DL})$ represents the mapping function of the PFN from $\widetilde{\mathbf Y}_{\rm DL}[k]$ to ${\mathbf q}[k]$. Since we assume that different terrestrial users are mutually independent and identically distributed (i.i.d.), all the terrestrial users use the identical PFN.

\begin{figure*}[t]
	\vspace{-3.0mm}
	\begin{center} 
		\includegraphics[width = 1\columnwidth, keepaspectratio]{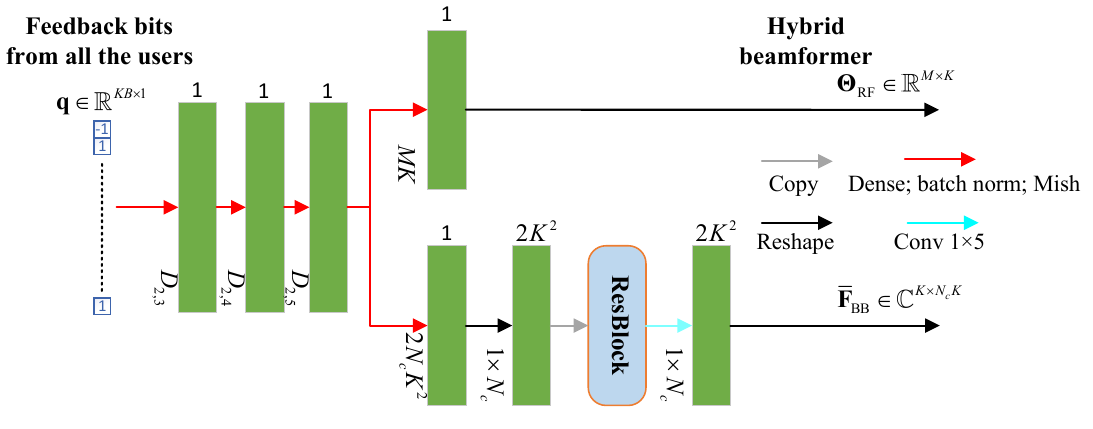}
	\end{center}
	\captionsetup{font = {footnotesize}, singlelinecheck = off, justification = raggedright, name = {Fig.}, labelsep = period}%
	\vspace{-3.0mm}
	\caption{Proposed FDD-HBFN for downlink multi-user broadband hybrid beamforming at the aerial BS.}
	\vspace{-3.0mm}
	\label{fig_FDDHB}
\end{figure*}

\subsection{Downlink Broadband Hybrid Beamforming for Multi-User}\label{S4.3}
In this paper, we assume that there is no feedback error from the terrestrial users to the aerial BS. Moreover, based on the feedback information, a multi-user hybrid broadband beamformer will be designed at the aerial BS. We will formulate this multi-user broadband hybrid beamforming process as an FDD-HBFN as shown in Fig.~\ref{fig_FDDHB}. The input of the FDD-HBFN is the feedback bit vector of all the terrestrial users $ {\mathbf q} = [{\mathbf q}[1]^T,{\mathbf q}[2]^T,\cdots,{\mathbf q}[{K}]^T]^T \in \mathbb{R}^{KB\times 1}$. The first four layers of the FDD-HBFN are four dense layers, and the numbers of neurons in these four dense layers are $ D_{2,3} $, $ D_{2,4} $, $ D_{2,5} $, and $ 2N_cK ^ 2 + MK $, respectively. The output of the fourth dense layer is divided into two parts. The first part (i.e., the upper branch in Fig.~\ref{fig_FDDHB}) with the $MK$-dimensional input will output the phase values ${\mathbf \Theta}_{\rm RF}\in \mathbb{R}^{M\times K}$ of the analog beamformer. The second part (i.e., the lower branch in Fig.~\ref{fig_FDDHB}) with the $2K^2N_c$-dimensional input will output the digital beamformer without power normalization. The above process can be expressed as 
\begin{align}\label{form16}
	\{ {\mathbf \Theta}_{\rm RF}, {\mathbf {\overline F}}_{{\rm BB}} \}=
	{\cal P}\left({\mathbf q};{\mathbf W}_{ T}^{\rm DL}\right),
\end{align}
where the mapping function of the FDD-HBFN is denoted by ${\cal P}(\cdot;{\mathbf W}_{ T}^{\rm DL})$, and ${\mathbf W}_{ T}^{\rm DL}$ is the set of the learnable weight parameters of the FDD-HBFN. Note that to normalize the transmit power, the process based on formula (\ref{form17}) and formula (\ref{form18}) will be performed to obtain the final analog beamformer  ${\mathbf F}_{\rm RF}$ and digital beamformer ${\mathbf F}_{{\rm BB}}[n]$ for $1\leq n\leq N_c$. 

In sum, the learnable parameters of the proposed E2E model in FDD mode include the phase matrix ${\mathbf \Phi}_{\rm DL}$ at the aerial BS, the learnable parameter set ${\mathbf W}_{R}^{\rm DL}$ of the PFN at the terrestrial users, and the learnable parameter set ${\mathbf W}_{T}^{\rm DL}$ of the FDD-HBFN at the aerial BS.

\subsection{Adapt E2E Neural Network to Quantized PSs Constraint via Transfer Learning \protect\footnote{In this subsection, we only consider the quantized PSs constraint in FDD systems. Since the quantized PSs constraint in FDD systems and TDD systems are identical, the proposed scheme can be directly applied to TDD systems.}}\label{S4.3}

\begin{figure*}[t]
	\vspace{-6.0mm}
	\begin{center} 
		\includegraphics[width = 1\columnwidth, keepaspectratio]{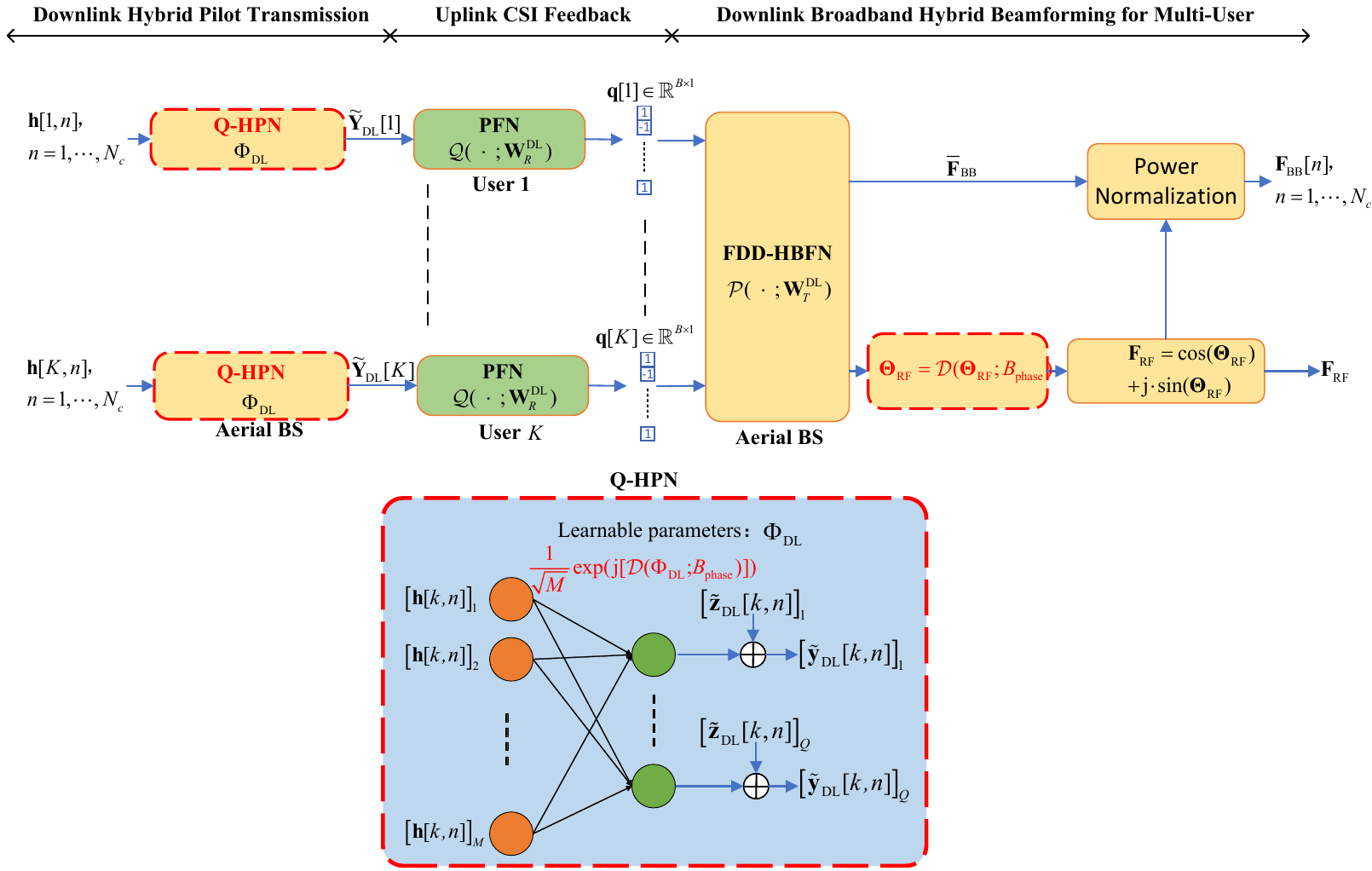}
	\end{center}
	\captionsetup{font = {footnotesize}, singlelinecheck = off, justification = raggedright, name = {Fig.}, labelsep = period}%
	\vspace{-6.0mm}
	\caption{Proposed DL-based scheme for FDD aerial massive MIMO-OFDM systems with quantized PSs constraint, where the red words indicate that the quantized PSs constraint
	is introduced into the corresponding modules.}
	\vspace{-6.0mm}
	\label{FIG_sys3}
\end{figure*}

In the above system, we have assumed the ideal infinite-resolution PSs, which are unrealistic in actual systems. Therefore, we design a novel E2E neural network as shown in Fig.~\ref{FIG_sys3} tailored for low-resolution PSs, and propose a new training strategy to adapt the proposed E2E neural network to quantized PSs constraint.

As for the hybrid pilot training, we propose a quantized-HPN (Q-HPN) as shown in Fig.~\ref{FIG_sys3} for generating the training signals under the constraint of quantized PSs, 
 where the trainable weight parameters are ${\mathbf \Phi}_{\rm DL}$, and the corresponding quantized mapping function is denoted as ${\cal D}(\cdot;B_{\rm phase})$.
For ${\cal D}(\cdot;B_{\rm phase})$,  $2^{B_{\rm phase}}$ quantized phase values are evenly divided from the range of $[0, 2\pi]$, and ${\cal D}(\cdot;B_{\rm phase})$ outputs the quantized phase value closest to the input phase value.
The quantization of network weight parameters was originally proposed in the binary neural network (BNN) \cite{BNN} for network compression. Here we use this method to accommodate the quantized PSs constraint. 
As for the downlink broadband hybrid beamforming, we adopt the same method to meet the quantized PSs constraint of the analog beamformer. 
To transfer the gradient in the training stage, the gradient of the quantization function is set to 1.

The discrete phase constraint makes it difficult to converge if we directly train the E2E neural network in Fig.~\ref{FIG_sys3}. 
To this end, {\color{black}we adopt a training strategy as described in Algorithm \ref{Alg1} based on the transfer learning method \cite{transfer}}, which is to apply the model to new and relevant tasks on the premise of obtaining some additional data or existing models. 
Specifically, the first step is pre-training. We firstly train an E2E neural network as the pre-trained model by assuming the ideal infinite-resolution PSs. The second step is to transfer the network weight parameters of the pre-trained model to the E2E neural network in Fig.~\ref{FIG_sys3}. Note that although the PSs' structures of these two networks are different, the dimensions of their trainable weight parameters are the same and have many common features, which can benefit the training of the E2E neural network in Fig.~\ref{FIG_sys3}.
The last step is fine-tuning. After transferring the parameters, we will further train the E2E neural network in Fig.~\ref{FIG_sys3} with a low learning rate to mitigate the performance loss caused by the phase quantization error.



%

\subsection{Training Strategy for Robust Generalizability}

{\color{black}In this paper, we consider using unsupervised learning to train the proposed E2E neural network in both TDD and FDD systems, where the negative sum rate is used as the loss function, which can be expressed as
	\begin{equation}\label{form32}
		{\cal{L}}_{\rm E2E} = -R = -\dfrac{1}{N_{c}} \sum\limits_{k=1}^K\sum\limits_{n=1}^{N_c}  
		{\rm log}_{2}\left( 1+\dfrac{\big| {\mathbf h}^H[{k,n}]{\mathbf F}_{\rm RF}{\mathbf f}_{\rm BB}[{k,n}] \big| ^{2}}{\sum_{k'\neq k}\big| {\mathbf h}^{H}[{k,n}]{\mathbf F}_{\rm RF}{\mathbf f}_{\rm BB}[{k',n}] \big| ^{2}+\sigma_n^2 } \right).
	\end{equation}
	The sum rate can be maximized by minimizing the loss function ${\cal{L}}_{\rm E2E}$ of the proposed E2E neural network.}

When training the proposed E2E neural network, we have assumed a specific system configuration, which may degrade the performance when the assumed parameters mismatch the actual systems. 
 Therefore, the generalizability to adapt the model to different environmental parameters is essential. In the following, we will discuss the generalizability to the number of channel paths $L_p$, the number of simultaneously served terristrial users $K$, the resolution of the PSs $B_{\rm phase}$, and SNR, respectively.

 \SetAlgoNoLine
 \SetAlCapFnt{\normalsize}
 \SetAlCapNameFnt{\normalsize}\
 \begin{algorithm}[!t]
 	
 	\color{black}
 	\caption{Training Strategy for Quantized PSs Constraint}\label{Algorithm:1}
 	\begin{algorithmic}[1]
 		\label{Alg1}
 		\STATE \textbf{\textbf{\%} Pre-training}
 		\STATE \textbf{Initialize} the phase matrix ${\mathbf \Phi}_{\rm DL,p}$, the learnable parameter set ${\mathbf W}_{R}^{\rm DL,p}$ of the PFN, and the learnable parameter set ${\mathbf W}_{T}^{\rm DL,p}$ of the FDD-HBFN for pre-trained model assuming the ideal infinite-resolution PSs;
 		\FOR {$l=1$ to $num\_epochs$}
 		\STATE \textbf{Update} ${\mathbf \Phi}_{\rm DL,p}$, ${\mathbf W}_{R}^{\rm DL,p}$, and ${\mathbf W}_{T}^{\rm DL,p}$ by minimizing the loss function ${\cal{L}}_{\rm E2E}$ in formula (\ref{form32});
 		\ENDFOR
 		
 		\STATE
 		\STATE \textbf{\textbf{\%} Parameter transfer}
 		\STATE \textbf{Initialize} ${\mathbf \Phi}_{\rm DL}$, ${\mathbf W}_{R}^{\rm DL}$, and ${\mathbf W}_{T}^{\rm DL}$ for the E2E neural network under quantized PSs constraint in Fig.~9 based on the pre-trained ${\mathbf \Phi}_{\rm DL,p}$, ${\mathbf W}_{R}^{\rm DL,p}$, and ${\mathbf W}_{T}^{\rm DL,p}$.
 		
 		\STATE
 		\STATE \textbf{\textbf{\%} Fine-tuning}
 		\FOR {$l=1$ to $num\_epochs$}
 		\STATE \textbf{Update} ${\mathbf \Phi}_{\rm DL}$, ${\mathbf W}_{R}^{\rm DL}$, and ${\mathbf W}_{T}^{\rm DL}$ by minimizing the loss function ${\cal{L}}_{\rm E2E}$ in formula (\ref{form32});
 		\ENDFOR
 		
 	\end{algorithmic}
 \end{algorithm}

 \subsubsection{Generalizability to different numbers of channel paths $L_p$}
 
 The CSI to be estimated and fed back is highly related to the number of channel paths $L_p$, which inevitably affects the channel estimation and feedback performance when the overhead is limited. 
  On the other hand, $L_p$ usually varies, which would degrade the performance if the network is trained under a fixed $L_p$. 
 Therefore, to enhance the generalizability of the proposed E2E neural network, we propose to train the network using the samples with different $L_p$.

\subsubsection{Generalizability to different numbers of terrestrial users $K$}
Since the channels of the different terrestrial users are typically i.i.d., all the terrestrial users can share the same parameters of the PFN, indicating that retraining the PFN is not required even though the number of terrestrial users $K$ changes. Therefore, we only need to train a PFN in the case of $K=1$, and the same PFN can be adopted by users with different $K$. However, for the FDD-HBFN at the aerial BS, since the dimension of the output beamforming matrix changes with the change of $K$, the FDD-HBFN requires to be retrained under different $K$.

\subsubsection{Generalizability to different resolutions of the PSs $B_{\rm phase}$}
With the help of the transfer learning method, we only need to fine-tune the E2E neural network under different $B_{\rm phase}$  based on the pre-trained network that assumes the ideal infinite resolution PSs.

\subsubsection{Generalizability to different SNRs}
In the case of different SNRs, the anti-noise robustness of the network can be enhanced 
by using the noisy training samples with different noise power.

\section{Numerical Results}\label{S6}
In this section, we investigate the performance of the proposed DL-based E2E approach with extensive simulations, where the involved baseline algorithms, training data, and training settings will be provided in detail.
\vspace{-2.0mm} 
\subsection{Baseline Algorithms}\label{S5.1}
\subsubsection{{\color{black}{\textbf{SS-HB}}\protect\footnote{SS-HB and TS-HB algorithms are originally designed for narrowband MIMO systems, but they can be easily extended to broadband MIMO-OFDM systems. In narrowband MIMO systems, SS-HB algorithm greedily selects analog beamforming vector from the codebook according to the correlation between the codebook and optimal fully-digital beamforming matrix. As for broadband MIMO-OFDM systems, since AoDs on all subcarriers are identical, the analog beamformer can be greedily selected from the codebook according to the sum of correlations between the codebook and optimal fully-digital beamforming matrix on all subcarriers, as the baseline used in \cite{Sun_TWC}. Similarly, the TS-HB algorithm can also be extended to broadband MIMO-OFDM systems.}/\textbf{TS-HB}/\textbf{MO-HB}\protect\footnote{\color{black}Manifold optimization based hybrid beamforming (MO-HB) is one of the alternate minimization hybrid beamforming schemes proposed in \cite{xiang_JSTISP}, and it  has the highest computational complexity and the best throughput performance among all schemes mentioned in \cite{xiang_JSTISP}. }/\textbf{PCA-HB}}  with perfect CSI at the aerial BS}\label{S5.1.1}
{\color{black}Consider the perfect CSI at the aerial BS, channel estimation and feedback are not required, and the aerial BS performs hybrid beamforming using SS-HB \cite{OMP_HP1}, TS-HB \cite{TSHP}, MO-HB \cite{xiang_JSTISP}, or PCA-HB \cite{Sun_TWC}.}

\subsubsection{{\color{black}{\textbf{SS-HB}}/\textbf{TS-HB}/\textbf{MO-HB}/\textbf{PCA-HB}} with the estimated CSI perfectly fed back to the aerial BS}\label{S5.1.2}
For TDD systems, we consider that each terrestrial user sends uplink pilot signals to the aerial BS, and the aerial BS utilizes conventional SW-OMP-based channel estimator \cite{CS} to estimate the CSI for the following downlink multi-user broadband hybrid beamforming. For FDD systems, we consider that the aerial BS broadcasts the downlink pilot signals to all terrestrial users, and each terrestrial user utilizes the SW-OMP-based channel estimator \cite{CS} for downlink CSI estimation, and the estimated CSI is assumed to be perfectly fed back to the aerial BS for the following downlinkmulti-user broadband hybrid beamforming.

\subsubsection{{\color{black}{\textbf{MO-HB}}/\textbf{PCA-HB}} with the estimated CSI and limited feedback for FDD systems}\label{S5.1.3}
In this case, the number of feedback bits is limited. Specifically, each terrestrial user first utilizes the conventional SW-OMP-based channel estimator to estimate the sparse channel parameters, which are then quantized according to the Lloyd max algorithm \cite{lloyd}. Assume the channel between each terrestrial user and the aerial BS to have $L_p$ dominated paths, the parameters required to be fed back are $\left\{ \Re({ \alpha}_{l,k}),\Im({ \alpha}_{l,k},) ,\theta_{l,k},\phi_{l,k},\tau_{l,k}\right\}$  for $1\leq l \leq L_p$, i.e., $5L_p$ parameters.
 We evenly allocate the total feedback bits to each parameter and utilize Lloyd max algorithm for quantization. According to the quantized parameters, the aerial BS reconstructs the CSI for performing hybrid beamforming. 
 
 {\color{black}
 \subsubsection{DL-based schemes \cite{E2EHB2,E2EHB4} \protect\footnote{ {\color{black}DL-based scheme \cite{E2EHB2} are originally designed for narrowband TDD massive MIMO systems, but we can extend it to broadband TDD massive MIMO-OFDM systems by summing the loss functions associated with  all subcarriers. Besides, DL-based schemes \cite{E2EHB2,E2EHB4} divide CSI acquisition and beamforming into digital and analog stages, so we take the sum of the pilot overheads of these two stages as the total pilot overhead. Besides, since \cite{E2EHB2} considers the quantization of the PSs' phase values, we will use it as a baseline method in TDD scenario with low-resolution PSs. }}}\label{S5.1.4}
 In this case, the BS uses DL-based schemes \cite{E2EHB2,E2EHB4} to perform hybrid beamforming according to the received pilot signals.
}


\subsection{Channel Samples for Network Training }\label{S5.2}
The training channel sample set is generated according to the sparse multipath channel model of aerial MIMO-OFDM systems, as described in Section \ref{S2}. Specifically, we consider the aerial BS is equipped with an $8\times 8$ UPA (i.e., $M=64$) having the half-wavelength antenna spacing, and the number of subcarriers $N_c$ is 32. 
The complex gains of different channel paths follow the i.i.d. complex Gaussian distribution, i.e., $\alpha_{l,k} \sim {\cal CN}\left( {0},1 \right) $, and the azimuth and zenith angles observed at the aerial BS follow the i.i.d. uniform distribution, i.e., $\theta_{l,k} \sim {\cal U} (-\pi/2,\pi/2)$, $\phi_{l,k} \sim {\cal U}  (-\pi/2,\pi/2)$.
Moreover, since the receiver has AWGN in both the channel estimation stage and downlink beamforming stage, we set the SNR to $ {\rm 10log_{10}}\sqrt{\frac{P_t}{N_c\sigma_n^2}}={\rm 10 }$ dB in the training stage. Note that we set the number of paths $L_p$ and the number of terrestrial users $K$ to 2, which is relatively small, so that the network training can be speed up to verify the proposed schemes. In the following subsections, we will adjust these parameters and observe the generalization performance of the proposed schemes under different system parameters.

\subsection{Training Settings}\label{S5.3}

We use the open-source deep learning library PyTorch to train and evaluate the proposed neural networks. We divide the data set into the training set, validation set, and test set, which contain 204800, 20480, and 20480 samples, respectively. So these three data sets are mutually exclusive. In the network training stage, the negative sum rate is chosen as the loss function and the Adam optimizer is adopted to update the network weight parameters. We set the batch size of the training set to 1024, and the training set contains a total of 200 batches. Epochs are set to 200. We set the initial learning rate to $10^{-3}$ and multiply the learning rate by 0.3 at the 100th and 150th epochs to further improve the performance in the later stage of training. 
In the training process, we use the early stopping strategy to monitor the sum rate based on the validation set, retain a set of network weight parameters having the best generalization performance under the validation set, and stop training when the sum rate based on the validation set has not increased for a relatively long time. 
 
In TDD systems, the numbers of neurons in the hidden dense layers of the proposed E2E neural network are set to $D_{1,1}=1024$, $D_{1,2}=512$, $D_{1,3}=128$, $D_{1,4}=2048$, $D_{1,5}=1024$, and $D_{1,6}=512$, respectively. 
In FDD systems, the numbers of neurons in the hidden dense layers of the proposed E2E neural network are set to $D_{2,1}=1024$, $D_{2,2}=512$, $D_{2,3}=2048$, $D_{2,4}=1024$, and $D_{2,5}=512$, respectively. 
The numbers of feature maps in the convolutional layers of the ResBlock are set to $C_1=256$ and $C_2=512$, respectively.

\begin{figure}[t]
	\vspace{-13.0mm}
	\captionsetup{font={footnotesize, color = {black}}, name = {Fig.}, labelsep = period} 
	\captionsetup[subfigure]{singlelinecheck = on, justification = raggedright, font={footnotesize}}
	\centering
	\includegraphics[scale=0.7]{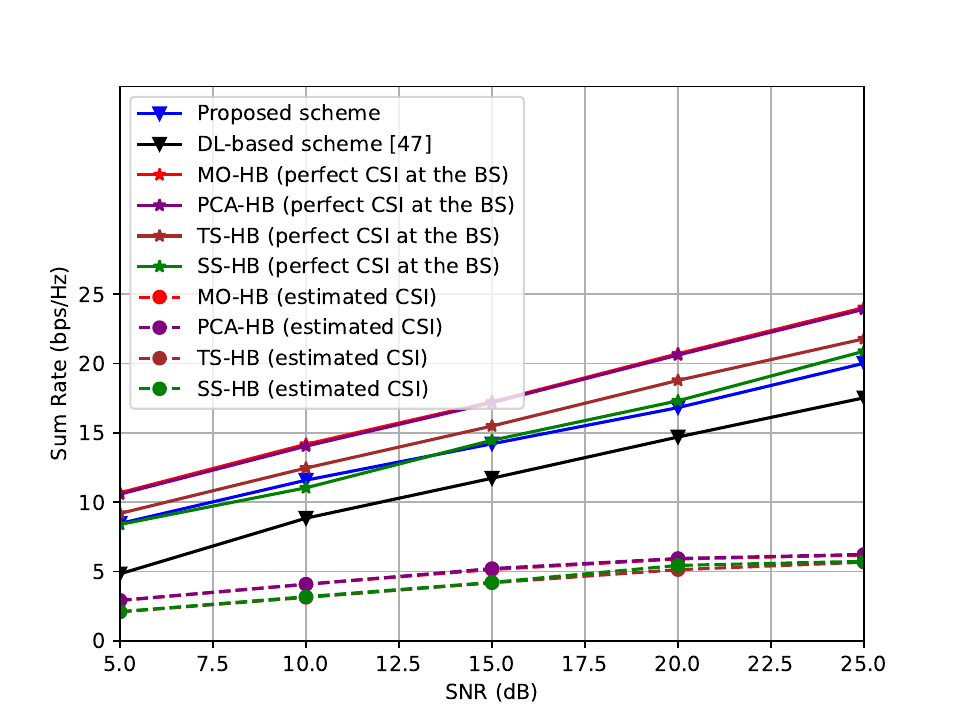} 
	\vspace{-2.0mm}
	\caption{Sum rate achieved by different schemes versus SNRs for TDD aerial massive MIMO-OFDM systems ($K=2$, $L_p=2$, $Q=2$). }\label{fig11}
	\vspace{-6.0mm}
\end{figure} 

\subsection{Performance in TDD Systems}\label{S5.4}
Fig.~\ref{fig11} shows the sum rate achieved by different schemes in TDD systems, where the number of pilot OFDM symbols is $Q=2$. 
{\color{black}It can be observed that MO-HB and PCA-HB using perfect CSI have the best performance, while the performance of the proposed DL-based E2E scheme can approach them and considerably outperforms the schemes based on the estimated CSI. }
Note that as the number of pilot OFDM symbols is $Q=2$, the effective dimension of channel estimation measurements observed at the aerial BS is $QK=4$, which is far smaller than the number of antennas $M=64$. In this case, the traditional CS-based explicit channel estimation schemes \cite{CS} can not well extract the sufficient CSI from the limited measurements. By contrast, the proposed DL-based E2E scheme can directly output the hybrid beamformer design from the received pilot signals  according to the mapping relationship, which is learned based on a large number of training samples, so that the explicit CSI estimation can be avoided. Therefore, the proposed scheme can achieve a better performance when the explicit CSI is difficult to be estimated due to the limited pilot overhead. This advantage is particularly suitable for the aerial massive MIMO-OFDM systems with limited channel coherence time.
{\color{black}Besides, note that the proposed scheme outperforms other DL-based technologies such as \cite{E2EHB4}, 
which  separately designs the analog and digital channel sensing and beamforming modules. This observation further indicates the superior performance of the proposed scheme that jointly models the key transmission modules as an E2E neural network.}

\subsection{Performance in FDD Systems}\label{S5.4}

Fig.~\ref{fig34} provides the sum rate comparison of different schemes versus the numbers of feedback bits $B$ for each terrestrial user, where $Q=8$ and $Q=4$ are investigated in Fig.~\ref{fig2} and Fig.~\ref{fig3}, respectively. 
{\color{black}It can be observed that the proposed scheme has a performance loss of no more than 3 bps/Hz compared to the MO-HB and PCA-HB with perfect CSI at the aerial BS even when the feedback overhead is limited to only 16 bits, while the latter one suffers from very high feedback overhead, which would lead to the large feedback latency due to the limited feedback link capacity. }This observation indicates the superior performance of the proposed E2E neural network approach for FDD aerial massive MIMO-OFDM systems in the case of very low CSI feedback overhead. 

 {\color{black}
\begin{figure*}[t]
	\vspace{-13.5mm}
	\captionsetup{font={footnotesize, color = {black}}, name = {Fig.}, labelsep = period} 
	\captionsetup[subfigure]{singlelinecheck = on, justification = raggedright, font={footnotesize}}
	\centering
	\hspace{-3.0mm}
	\subfloat[]{
		\label{fig2}
		\begin{minipage}[t]{0.48\linewidth}
			\centering
			\includegraphics[scale=0.49]{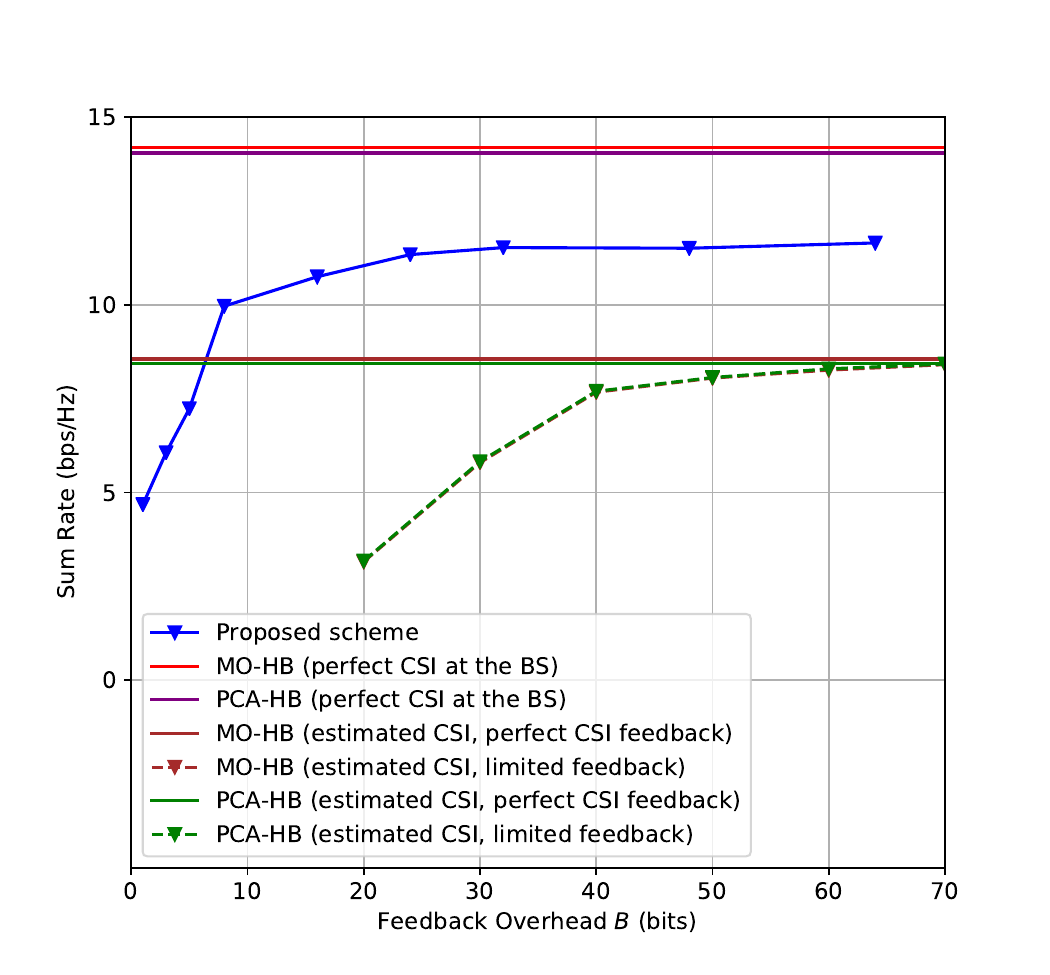}
		\end{minipage}
	}
	\hfill
	\subfloat[]{
		\label{fig3}
		\begin{minipage}[t]{0.48\linewidth}
			\centering
			\includegraphics[scale=0.49]{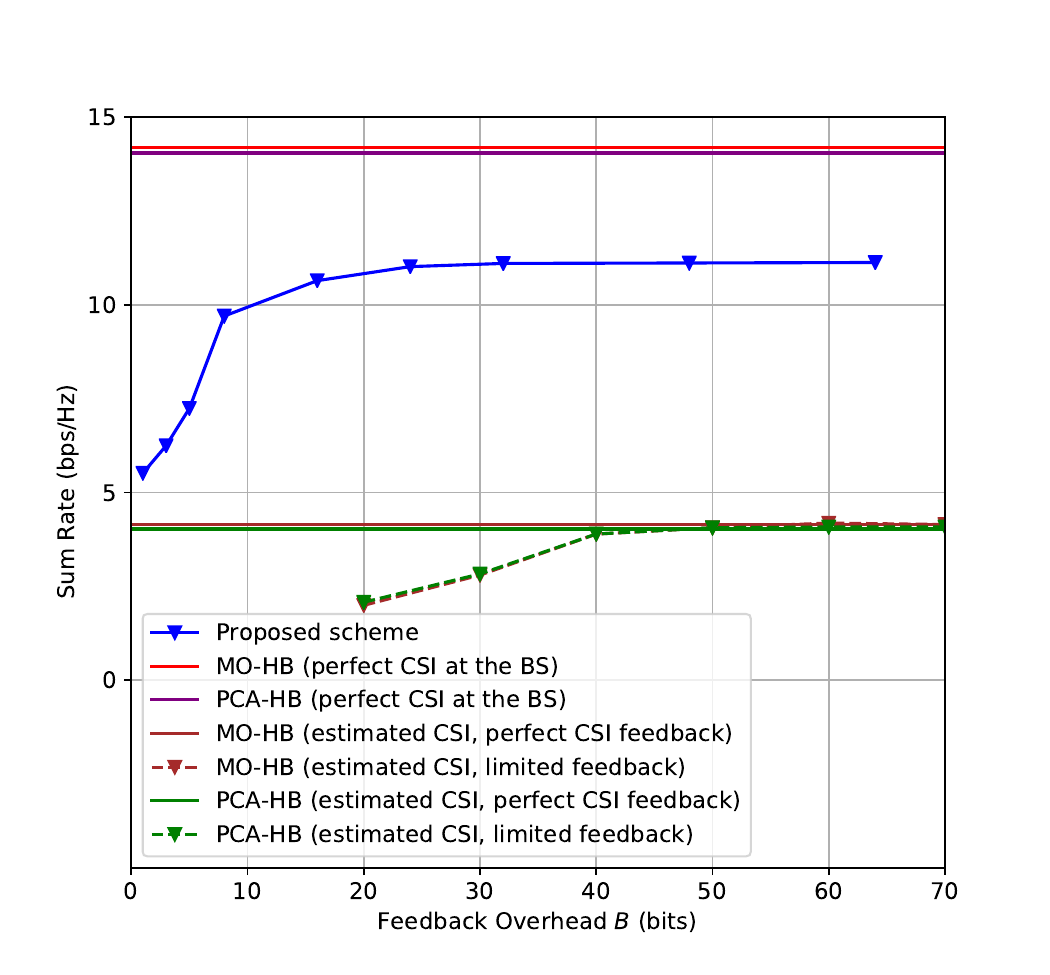}
		\end{minipage}
	}
	\vspace{-3.0mm}
	\caption{Sum rate achieved by different schemes versus the numbers of feedback bits $B$ for FDD aerial massive MIMO-OFDM systems ($K=2$, $L_p=2$, SNR = 10 dB): (a) $Q = 8$; (b) $Q = 4$.}
	\label{fig34}
	\vspace{-5.0mm}
\end{figure*}
}

%

On the other hand, we can observe that, when the number of pilot OFDM symbols $Q$ reduces from 8 to 4, the proposed scheme has no significant performance loss. By contrast, either the MO-HB or the PCA-HB with the estimated CSI and limited feedback suffers from the severe performance loss. Hence, the superiority of the proposed scheme in the case of limited channel estimation and feedback overhead is self-evident.


Fig.~\ref{fig4} compares the sum rate performance of different schemes versus SNRs, where we consider the number of pilot OFDM symbol is $Q=8$ and the number of feedback bits is $B=30$. It can be observed that the proposed scheme is significantly better than the traditional schemes with limited channel estimation and feedback overhead, which verify the robustness of the proposed E2E neural network approach to different SNRs.

Fig.~\ref{fig5} compares the convergence process of the proposed scheme with different numbers of feedback bits, where the number of pilot OFDM symbols is $Q=8$ and the number of terrestrial users is $K=2$. In simulations, we reduce the learning rate from $10^{-3}$ to $3\times 10^{-4}$ at the 100-th epoch. It can be seen that, to achieve the good performance, the 
proposed scheme requires around 50 epochs. In the next subsection, we will further discuss the generalizability of the proposed scheme and prove that the proposed network can effectively work when the number of mutlipath components $L_p$ and the number of simultaneously served terrestrial users $K$ become large.

 {\color{black}
\begin{figure*}[t]
	\vspace{-9mm}
	\captionsetup{font={footnotesize}, singlelinecheck = off, justification = raggedright,name={Fig.},labelsep=period}
	\centering
	\centering
	\begin{minipage}[t]{0.49\linewidth}
		\centering
		\includegraphics[scale=0.49]{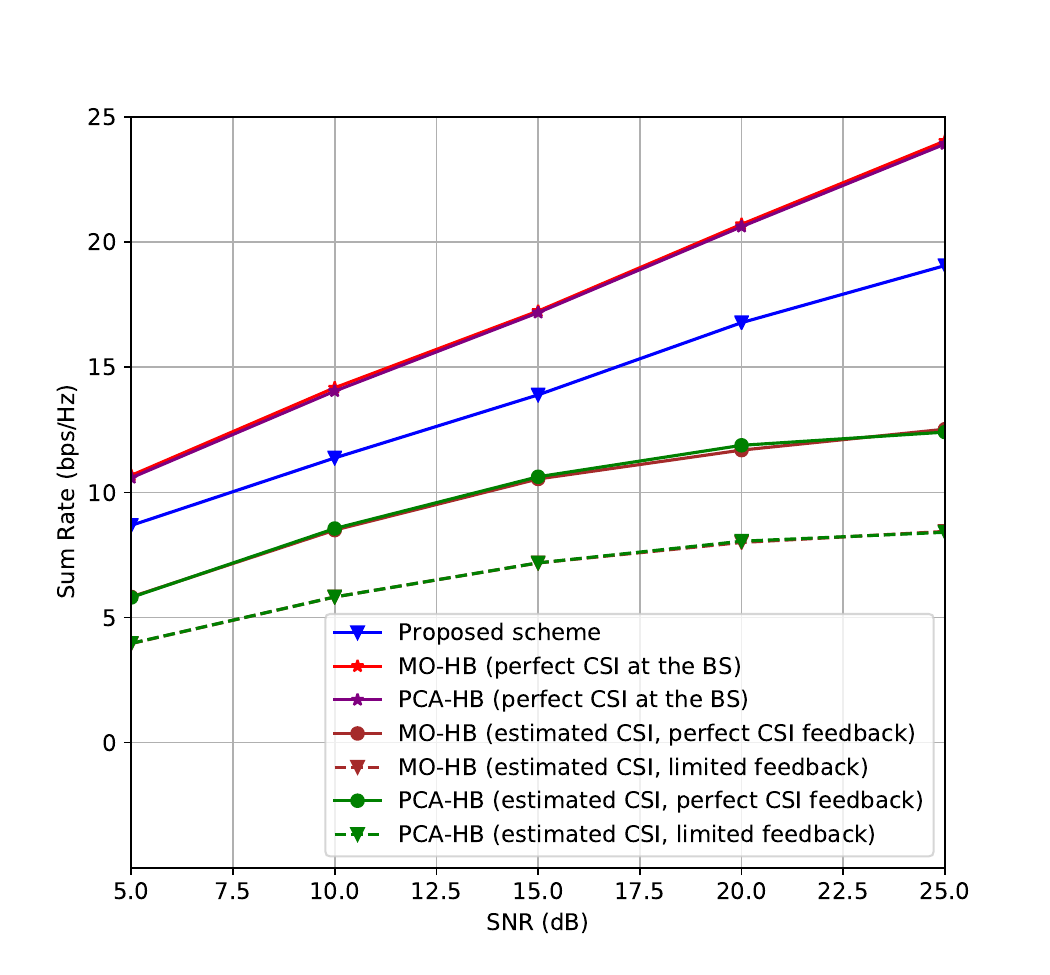}
		\vspace{-8.0mm}
		\captionsetup{font={footnotesize, color = {black}}, singlelinecheck = off, justification = raggedright,name={Fig.},labelsep=period}
		\caption{Sum rate achieved by different schemes versus SNRs for FDD aerial massive MIMO-OFDM systems ($Q=8$, $K=2$, $L_p=2$, $B=30$).}
		\label{fig4}
	\end{minipage}
	\hfill
	\begin{minipage}[t]{0.49\linewidth}
		\centering
		\includegraphics[scale=0.49]{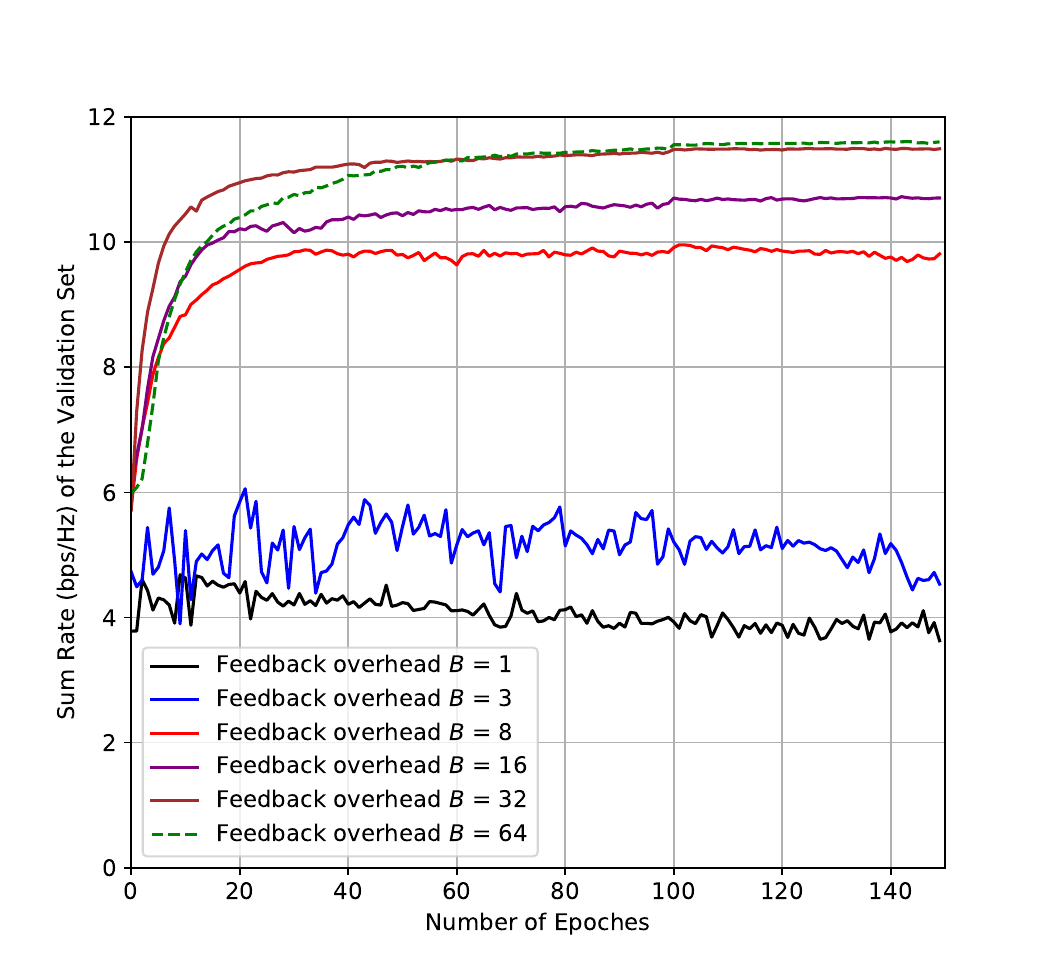}
		\vspace{-8.0mm}
		\caption{The convergence process of the proposed scheme in FDD mode ($Q=8$, $K=2$, $L_p=2$, SNR = 10 dB).}
		\label{fig5}
	\end{minipage}
	
	\vspace{-6.0mm}
\end{figure*}
}

\subsection{Generalizability}\label{S5.5}

In this subsection, we verify the generalizability of the proposed DL-based E2E scheme in FDD mode.  

Fig.~\ref{fig6} shows the sum rate achieved by different schemes versus the numbers of paths $L_p$, where the number of terrestrial users is $K=2$, the number of feedback bits is $B=30$, and the number of pilot OFDM symbols is  $Q=8$. In Fig.~\ref{fig6}, we investigate the performance of the proposed scheme with two different training strategies: the label `trained with $L_p=2$' indicates that the channel samples in the training set have a fixed number of $L_p=2$, and the label `trained with $L_p\in \cal{U}$[1,8]' indicates that $L_p$ of the channel samples in the training set follow the i.i.d. uniformly discrete distribution $\cal{U}$[1,8]. It can be seen that the second training strategy has a robust generalizability to $L_p$. Besides, the proposed scheme with both training strategies outperform either the MO-HB or PCA-HB with the estimated CSI and limited feedback overhead. 

 {\color{black}
\begin{figure*}[t]
	\vspace{-10mm}
	\captionsetup{font={footnotesize, color = {black}}, singlelinecheck = off, justification = raggedright,name={Fig.},labelsep=period}
	\centering
	\centering
	\begin{minipage}[t]{0.49\linewidth}
		\centering
		\includegraphics[scale=0.49]{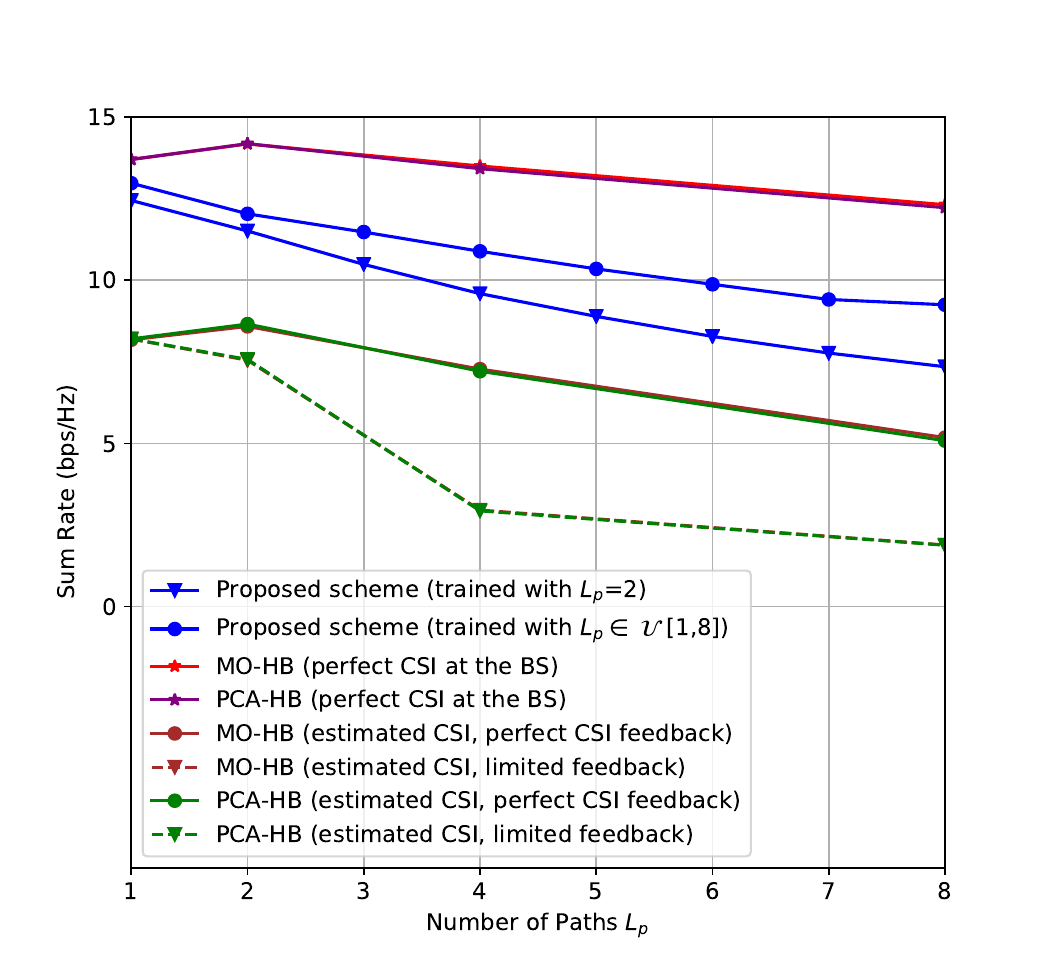}
		\vspace{-8.0mm}
		\caption{Sum rate achieved by different schemes versus $L_p$ for FDD aerial massive MIMO-OFDM systems ($Q=8$, $K=2$, $B=30$, SNR = 10 dB).}
		\label{fig6}
	\end{minipage}
	\hfill
	\begin{minipage}[t]{0.49\linewidth}
		\centering
		\includegraphics[scale=0.49]{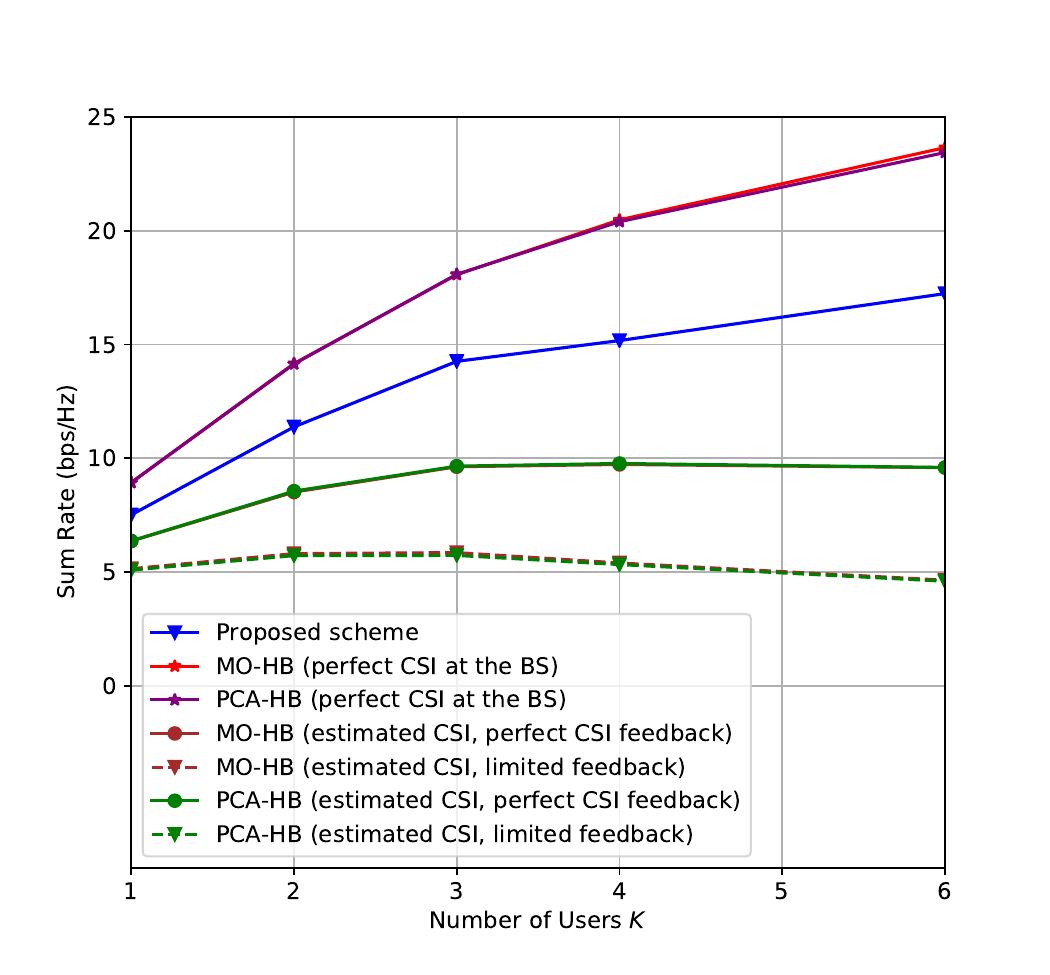}
		\vspace{-8.0mm}
		\caption{Sum rate achieved by different schemes versus $K$ for FDD aerial massive MIMO-OFDM systems ($Q=8$, $L_p=2$, $B=30$, SNR = 10 dB).}
		\label{fig7}
	\end{minipage}
	
	\vspace{-6.0mm}
\end{figure*}
\begin{figure*}[t]
	\vspace{-6mm}
	\captionsetup{font={footnotesize, color = {black}}, name = {Fig.}, labelsep = period} 
	\captionsetup[subfigure]{singlelinecheck = on, justification = raggedright, font={footnotesize}}
	\centering
	\hspace{-3.0mm}
	\subfloat[]{
		\label{FIG9(a)}
		\centering
		\begin{minipage}[t]{0.48\linewidth}
			\centering
			\includegraphics[scale=0.49]{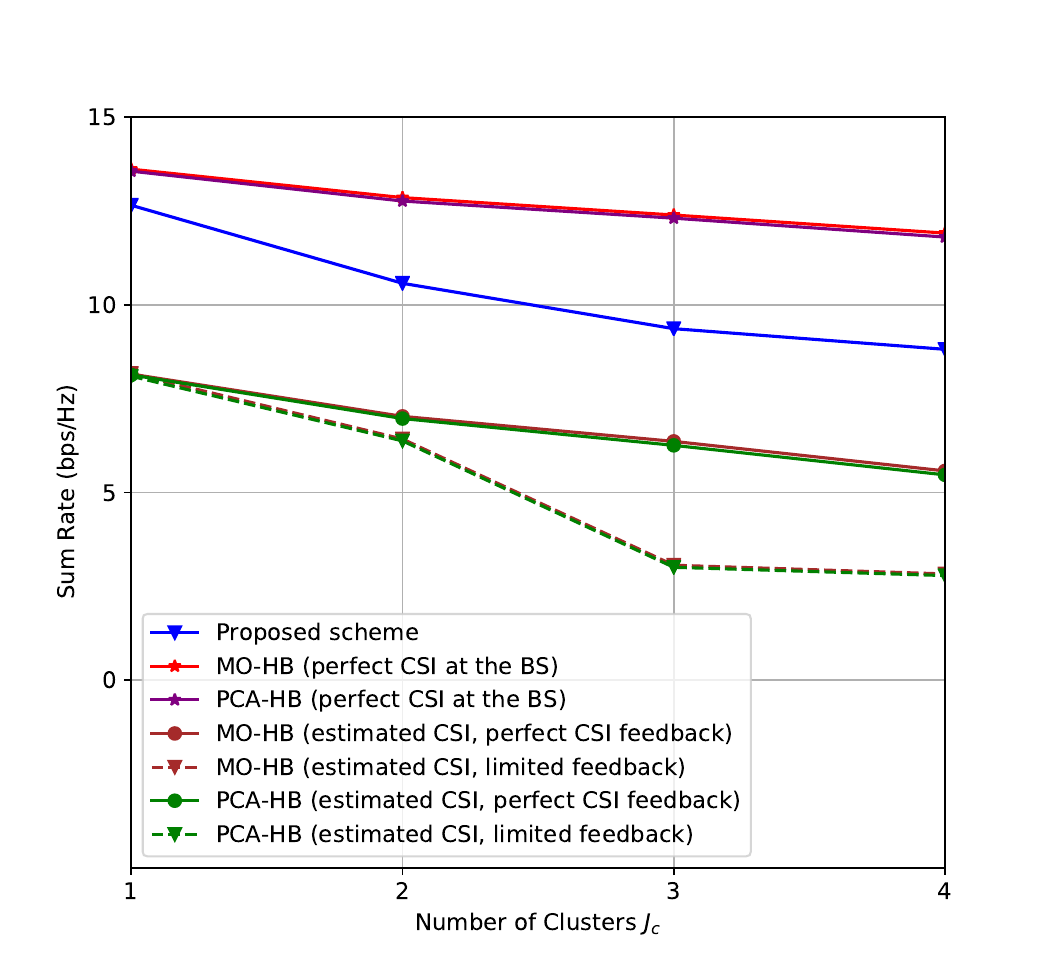}
			\vspace{-14.0mm}
		\end{minipage}
	}
	\hfill
	\subfloat[]{
		\label{FIG9(b)}
		\centering
		\begin{minipage}[t]{0.48\linewidth}
			\centering
			\includegraphics[scale=0.49]{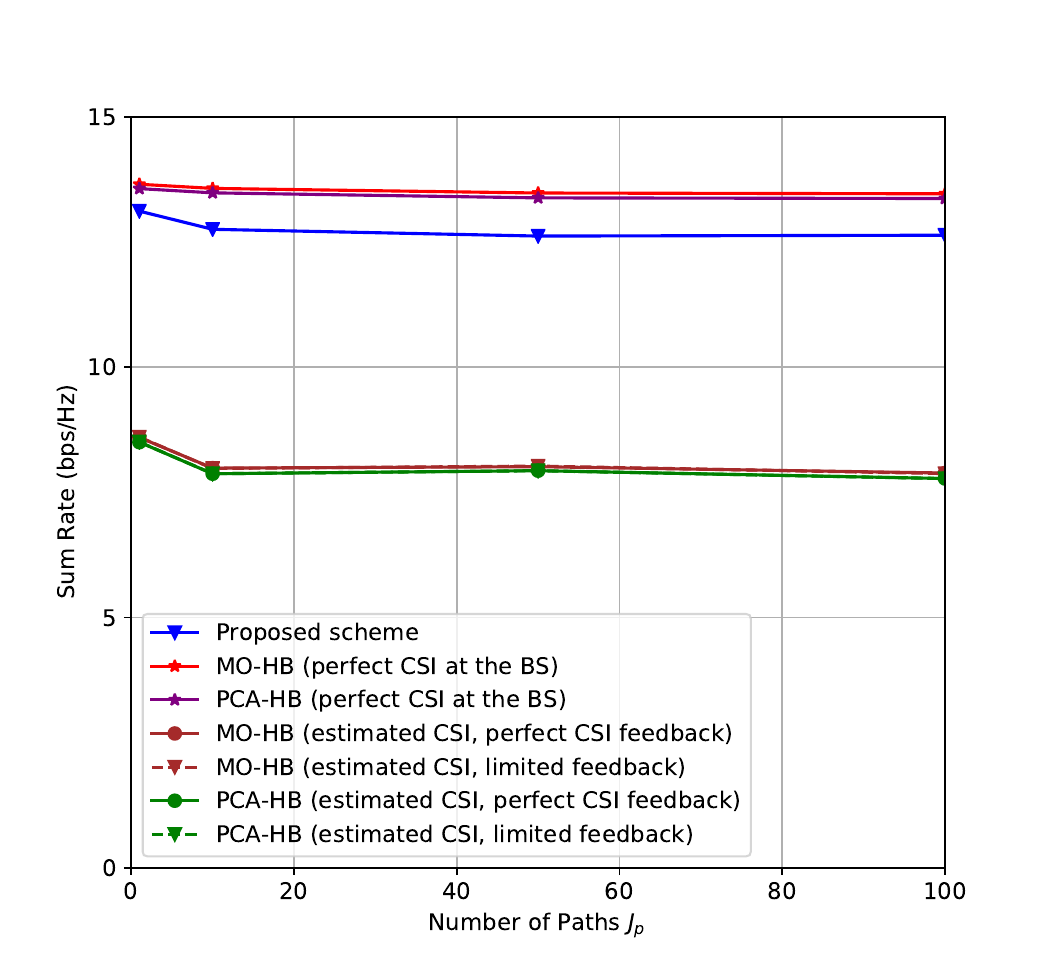}
			\vspace{-14.0mm}
		\end{minipage}
	}
	\vspace{-3.0mm}
	\caption{Sum rate achieved by different schemes for FDD aerial massive MIMO-OFDM systems with hybrid beamforming, where $B = 40$, SNR = 10 dB, and $K=2$:  (a)~Typical cluster sparse channel model; (b)~typical one-ring  channel model.}
	\label{FIG9}
	
	\vspace{-6.0mm}
\end{figure*}
}
Fig.~\ref{fig7} compares the sum rate achieved by different schemes versus the numbers of simultaneously served terrestrial user $K$ (also the numbers of RF chains) for FDD aerial massive MIMO-OFDM systems, where the number of pilot OFDM symbols is $Q=8$ and the number of feedback bits is $B=30$. Again, we can observe the robustness of the proposed scheme to different numbers of terrestrial users as well as the superiority over other schemes. 

\begin{figure}[t]
	\vspace{-13.0mm}
	\captionsetup{font={footnotesize, color = {black}}, name = {Fig.}, labelsep = period} 
	\captionsetup[subfigure]{singlelinecheck = on, justification = raggedright, font={footnotesize}}
	\centering
	\includegraphics[scale=0.7]{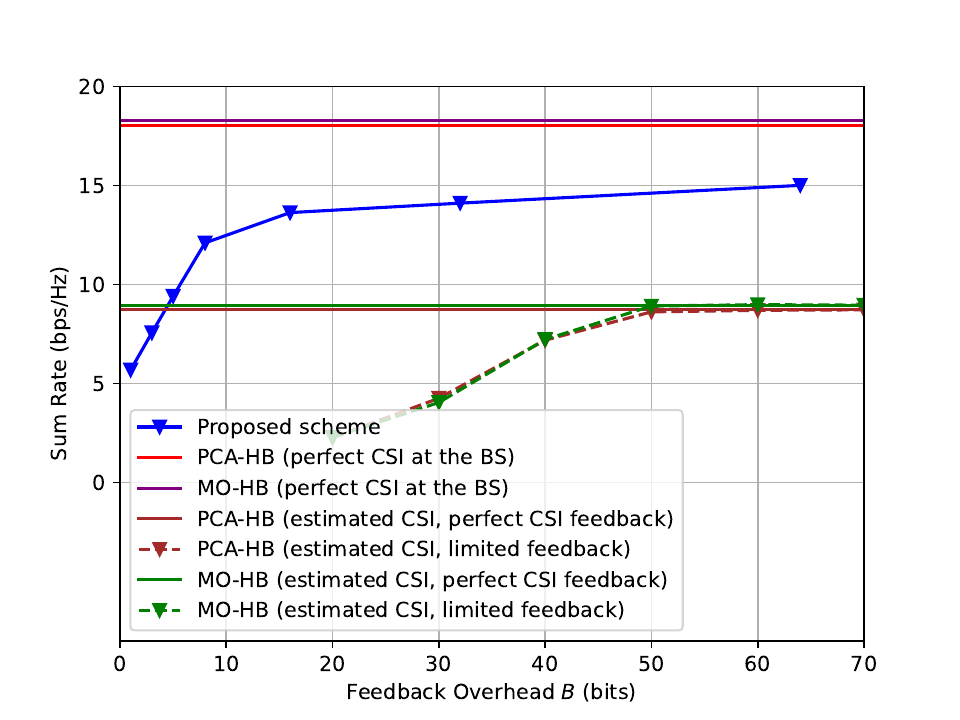} 
	\vspace{-2.0mm}
	\caption{Sum rate achieved by different schemes versus the numbers of feedback bits for FDD aerial massive MIMO-OFDM systems ($M = 16\times 16 = 256$, $K=2$, $L_p=2$, SNR = 10 dB, $Q = 8$). }\label{com_2}
	\vspace{-6.0mm}
\end{figure}

Fig.~\ref{FIG9(a)} and Fig.~\ref{FIG9(b)} investigate the sum rate achieved by different schemes under the cluster sparse channel model and one-ring channel model \cite{Ma_tvt}, respectively, where FDD aerial massive MIMO-OFDM systems with hybrid beamforming is considered, and $M=64$, $B=40$, and $K=2$ are adopted. Specifically, we consider there are $J_c$ scattering clusters between the aerial BS and each terrestrial user, and each scattering cluster has $J_p$ paths. Moreover, we consider the angle spread in each cluster is $\sigma_\theta=7.5^{\circ}$, and the delay spread in each cluster is $\sigma_\tau=T_s$. The downlink channel between the aerial BS and the $k$-th terrestrial user on the $n$-th subcarrier can be mathematically expressed as
\begin{align}\label{form20}
 {\mathbf h}[{k,n}] = \dfrac{1}{\sqrt{J_cJ_p}}\sum_{c=1}^{J_c}\sum_{p=1}^{J_p}\alpha_{k,c,p}{\mathbf a}_t(\theta_{k,c,p},\phi_{k,c,p})e^{-{\rm j}\frac{2\pi n\tau_{k,c,p}}{N_cT_s}},
\end{align}
where $\alpha_{k,c,p}\sim {\cal CN}\left( {0},1 \right)$ is the complex gain of the $p$-th path in the $c$-th scattering cluster, $\theta_{k,c,p}\sim {\cal U} \left(-\pi/2,\pi/2 \right)$ and $\phi_{k,c,p}\sim {\cal U} \left(-\pi/2,\pi/2 \right)$  are the azimuth and zenith AoDs of the $p$-th path in the $c$-th scattering cluster, respectively. For the cluster sparse channel model, we consider the number of clusters varies from 1 to 4, and the number of paths in each cluster is $J_p=10$. For the one-ring channel model, we consider the number of clusters is $J_c=1$, and the number of paths in each cluster varies from 1 to 100.  As shown in Fig.~\ref{FIG9}, the proposed scheme outperforms conventional schemes under both the cluster sparse channel model and the one-ring channel model.

{\color{black}The aforementioned simulations all consider that the number of BS antennas is $M=8\times8=64$. Fig.~\ref{com_2} further investigates the performance of the proposed scheme in the case where the BS is equipped with a larger MIMO array (i.e., the number of BS antennas is $M=16\times16=256$). It can be observed that the proposed E2E neural network can still achieve satisfactory performance when the number of BS antennas is $M=256$. }

 {\color{black}
 \begin{figure*}[t]
 	\vspace{-12mm}
 	\captionsetup{font={footnotesize, color = {black}}, name = {Fig.}, labelsep = period} 
 	\captionsetup[subfigure]{singlelinecheck = on, justification = raggedright, font={footnotesize}}
 	\centering
 	\hspace{-3.0mm}
 	\subfloat[]{
 		\label{fig8(a)}
 		\centering
 		\begin{minipage}[t]{0.48\linewidth}
 			\centering
 			\includegraphics[scale=0.49]{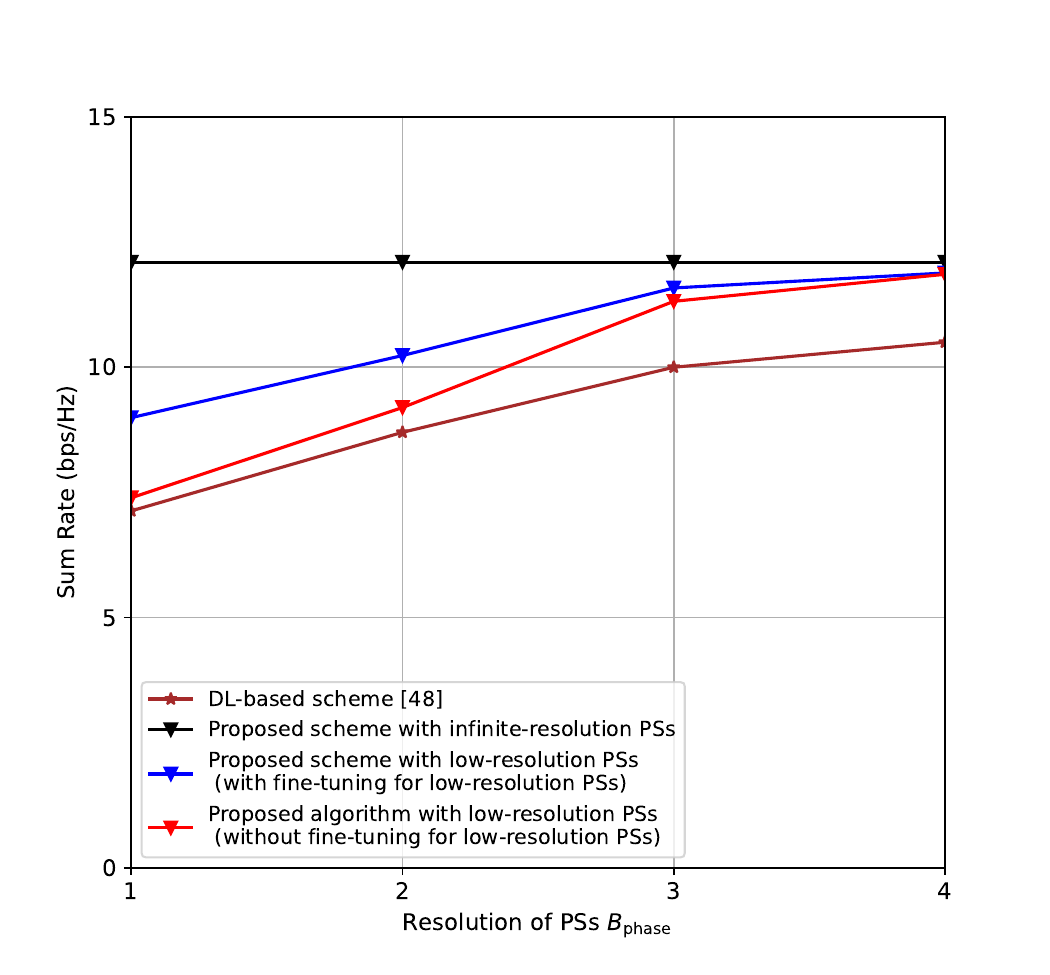}
 			\vspace{-14.0mm}
 		\end{minipage}
 	}
 	\hfill
 	\subfloat[]{
 		\label{fig8(b)}
 		\centering
 		\begin{minipage}[t]{0.48\linewidth}
 			\centering
 			\includegraphics[scale=0.49]{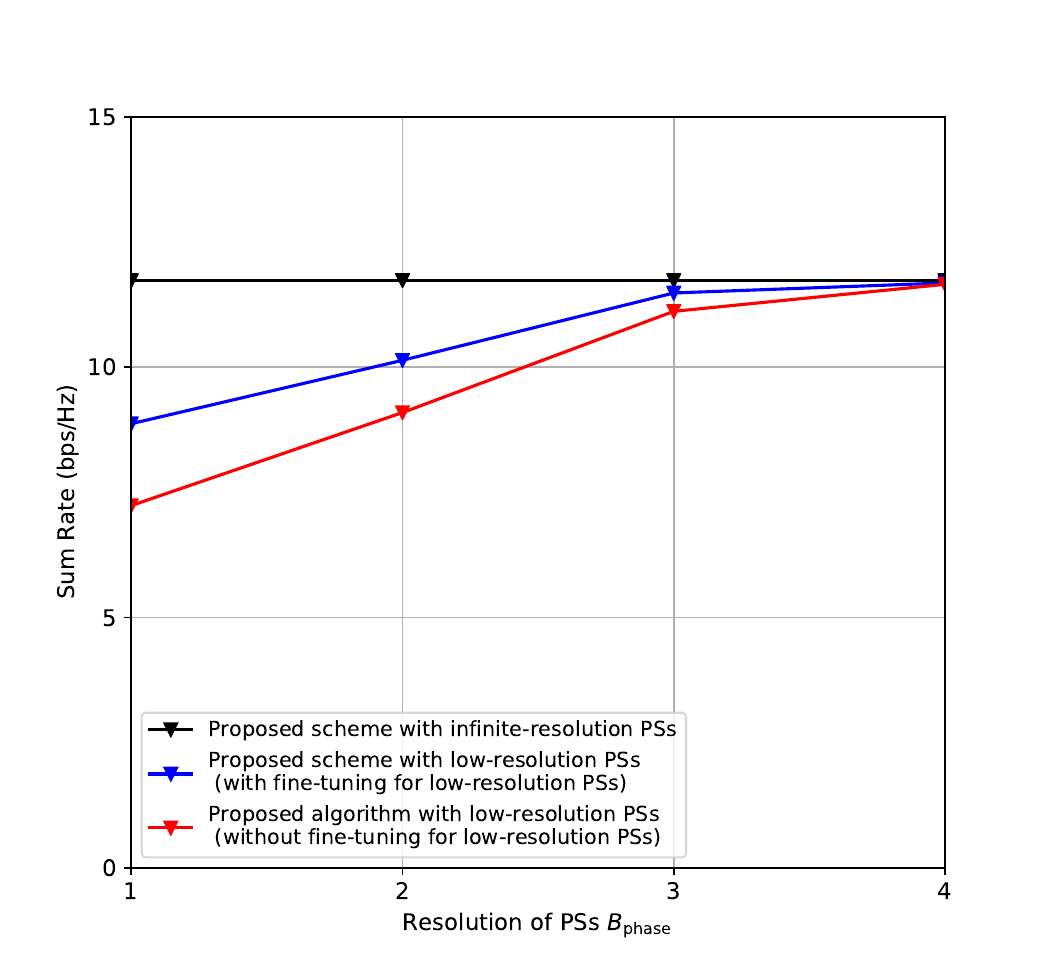}
 			\vspace{-14.0mm}
 		\end{minipage}
 	}
 	\vspace{-3.0mm}
 	\caption{Sum rate achieved by the proposed scheme under quantized PSs constraint:  (a)~TDD ($M=64$, $Q=4$, $K=2$, $L_p=2$, SNR = 10 dB); (b)~FDD ($M=64$, $Q=8$, $K=2$, $L_p=2$, $B=40$, SNR = 10 dB).}
 	\label{fig8}
 	\vspace{-6.0mm}
 \end{figure*}
 	
 }


\subsection{Effectiveness of the Proposed Method under Low-Resolution PSs}\label{S5.6}

{\color{black}Fig.~\ref{fig8(a)} and Fig.~\ref{fig8(b)} investigate the sum rate of the proposed scheme versus the resolutions of PSs (denoted by $B_{\rm phase}$) in TDD and FDD systems, respectively.} In simulations, the proposed scheme with infinite-resolution PSs provides the performance upper bound. It can be seen that the performance loss caused by phase quantization error can be reduced by adopting the proposed transfer learning strategy, i.e., 
fine-tuning the E2E neural network under the constraint of low-resolution PSs.
Fig.~\ref{fig8} also shows that the proposed scheme with the low-resolution PSs can achieve the performance close to that with infinite-resolution PSs, when the resolution of PSs is no less than 3. Even when PSs have only 1 bit or 2 bits resolution, the proposed E2E neural network can still work. 
{\color{black}Besides, the proposed scheme outperforms the existing DL-based scheme \cite{E2EHB2} in TDD scenario with low-resolution PSs as shown in Fig.~\ref{fig8(a)}, which indicates the superior performance of the proposed E2E neural network and training strategy.}
Therefore, the proposed scheme can be applied to the systems with low-resolution PSs for saving hardware cost and power consumption.

\begin{table*}[!tp]
	\centering  
	\color{black}
	\captionsetup{font={color = {black}}}
	\caption{Computational Complexity of Different Schemes.}  
	\label{table1}  
	\begin{tabular}{|c|l|}
		\hline
		\textbf{Schemes}                  & \multicolumn{1}{c|}{\textbf{Complexity}} \\ \hline
		Proposed                 & $\mathcal{O}\left(\sum_{i=1}^{N_{\mathrm{\rm de}}} d_{i-1} d_{i} + \beta KU N_{c} \sum_{i=1}^{N_{\rm c o}} n_{i-1} n_{i}\right)$          \\ \hline
		DL-based scheme \cite{E2EHB4} & $\mathcal{O}\left(\sum_{i=1}^{N_{\mathrm{de}}} d_{i-1} d_{i}\right)$          \\ \hline
		DL-based scheme \cite{E2EHB2} & $\mathcal{O}\left(\sum_{i=1}^{N_{\mathrm{\rm de}}} d_{i-1} d_{i} + \beta M K\sum_{i=1}^{N_{\rm co}} n_{i-1} n_{i}\right)$          \\ \hline
		SW-OMP                   & $\mathcal{O}\left(K U G^{2} N_{c} I\right)$          \\ \hline
		MO-HB                    & $\mathcal{O}\left( K M^{2} N_{c} I_{1} I_{2}\right)$          \\ \hline
		PCA-HB                   & $\mathcal{O}\left(K M^{2} N_{c}+K M N_{c}^{2}\right)$          \\ \hline
	\end{tabular}
\end{table*}

\begin{table*}[!tp]
	\centering  
	\color{black}
	\captionsetup{font={color = {black}}}
	\caption{Running Time of Different Schemes.}  
	\label{table2}  
	\begin{tabular}{|c|ccc|}
		\hline
		\textbf{Schemes}                  & \multicolumn{1}{c|}{\textbf{\begin{tabular}[c]{@{}c@{}}Channel estimation \\ and feedback\end{tabular}}} & \multicolumn{1}{c|}{\textbf{\begin{tabular}[c]{@{}c@{}}Hybrid\\ beamforming\end{tabular}}} & \multicolumn{1}{l|}{\textbf{Total time}} \\ \hline
		SW-OMP + MO-HB           & \multicolumn{1}{c|}{0.721 s in CPU}                                                                    & \multicolumn{1}{c|}{2.671 s in CPU}                                                      & 3.392 s   in CPU                       \\ \hline
		SW-OMP + PCA-HB          & \multicolumn{1}{c|}{0.721 s in CPU}                                                                    & \multicolumn{1}{c|}{0.027  s in CPU}                                                      & 0.748 s  in CPU                        \\ \hline
		DL-based scheme {\cite{E2EHB4}} & \multicolumn{3}{c|}{\begin{tabular}[c]{@{}c@{}}4.86 ms in GPU  or  16.8 ms in CPU\end{tabular}}                                                                                                                     \\ \hline
		DL-based scheme {\cite{E2EHB2}} & \multicolumn{3}{c|}{\begin{tabular}[c]{@{}c@{}}5.22 ms in GPU  or  28.2 ms in CPU\end{tabular}}                                                                                                                     \\ \hline
		Proposed                 & \multicolumn{3}{c|}{\begin{tabular}[c]{@{}c@{}}5.55 ms in GPU  or  31.8 ms in CPU\end{tabular}}                                                                                                                     \\ \hline
	\end{tabular}
\end{table*}

{\color{black}
\subsection{Computational Complexity Analysis }\label{S5.6}

This subsection investigates the computational complexity of different schemes. Since the offline training stage does not have a strict time limit, we mainly focus on the computational complexity of the online test stage. The details of computational complexity analysis of different schemes are as follows.
\begin{itemize}
	\item The main steps of the proposed scheme include: i) $N_{\rm de}$ dense layers with computational complexity $\mathcal{O}\left(\sum_{i=1}^{N_{\mathrm{de}}} d_{i-1} d_{i}\right)$, where $d_{i-1}$ and $d_{i}$ denote the input and output sizes of the $i$-th dense layer, respectively; ii) $N_{\rm co}$ convolutional layers with computational complexity $\mathcal{O}\left(\beta K U N_{c} \sum_{i=1}^{N_{c o}} n_{i-1} n_{i}\right)$, where $U$ ($U = QK$ in TDD systems or $U=Q$ in FDD systems) is the effective dimension of channel estimation measurements observed at the aerial BS, $\beta$ is the size of the convolutional filters, $n_{i-1}$ and $n_i$ denote the numbers of input and output feature maps of the $i$-th convolutional layer, respectively. Therefore, the computational complexity of the proposed scheme is $\mathcal{O}\left(\sum_{i=1}^{N_{\mathrm{de}}} d_{i-1} d_{i} + \beta K U N_{c} \sum_{i=1}^{N_{c o}} n_{i-1} n_{i}\right)$.
\end{itemize}
\begin{itemize}
	\item DL-based scheme \cite{E2EHB4} is mainly composed of dense layers and its computational complexity is $\mathcal{O}\left(\sum_{i=1}^{N_{\mathrm{de}}} d_{i-1} d_{i}\right)$.
\end{itemize}
\begin{itemize}
	\item The computational complexity of DL-based scheme \cite{E2EHB2} mainly comes from the dense layers and the convolutional layers, i.e., $\mathcal{O}\left(\sum_{i=1}^{N_{\mathrm{\rm de}}} d_{i-1} d_{i} + \beta M K\sum_{i=1}^{N_{\rm co}} n_{i-1} n_{i}\right)$.
\end{itemize}
\begin{itemize}
	\item As for SW-OMP channel estimation algorithm \cite{CS}, the computational complexity mainly comes from the correlation operation, i.e.,  $\mathcal{O}\left(K U G^{2} N_{c} I\right)$, where $G$ is the number of columns of the SW-OMP redundant dictionary matrix and $I$ is the number of iterations.
\end{itemize}
\begin{itemize}
	\item As for MO-HB algorithm, we denote the number of iterations as $I_1$. In each iteration of MO-HB algorithm, a conjugate gradient descent search is required to update the analog beamformer. Assuming that the number of conjugate gradient searches in each iteration is $I_2$, then the total computational complexity of MO-HB for $K$ users is $\mathcal{O}\left( K M^{2} N_{c} I_{1} I_{2}\right)$.
\end{itemize}
\begin{itemize}
	\item As for PCA-HB algorithm, the computational complexity mainly comes from SVD operation, i.e.,  $\mathcal{O}\left(K M^{2} N_{c}+K M N_{c}^{2}\right)$.
\end{itemize}

To intuitively observe the computational complexity of different schemes, TABLE~II shows the running time of different schemes. It can be observed that the running time of different DL-based schemes (i.e., \cite{E2EHB4,E2EHB2}, and the proposed scheme)  is similar and is significantly lower than that of model-based schemes (i.e., SW-OMP \cite{CS}, MO-HB \cite{xiang_JSTISP}, and PCA-HB \cite{Sun_TWC}). 

}

\section{Conclusion}\label{S6}

This paper proposes a DL-based E2E approach for aerial multi-user broadband hybrid beamforming. The proposed approach provides a unified hybrid beamforming framework for both TDD and FDD aerial massive MIMO-OFDM systems with implicit CSI. For TDD E2E neural network, we jointly optimize the uplink pilot combining and downlink hybrid beamforming, which respectively correspond to the HPN and the TDD-HBFN.  The HPN is used for training the learnable weights of the uplink pilot combining matrix, and the TDD-HBFN is used for extracting the implicit CSI from the received pilot signals to design hybrid beamforming. For FDD E2E neural network, we jointly optimize the downlink pilot transmission, uplink CSI feedback, and downlink hybrid beamforming, which correspond to the HPN, the PFN, and the FDD-HBFN, respectively. The HPN in FDD systems has the similar network structure as that in TDD systems, the PFN is used at the terrestrial users for extracting the implicit CSI from the received pilot signals and then compressing it into the binary bit vectors, and the FDD-HBFN is used at the aerial BS to directly map all terrestrial users' feedback bits to multi-user broadband hybrid beamformer. In addition, we introduce the transfer learning strategy to adapt the proposed scheme to quantized PSs constraint, so that the proposed scheme can mitigate the performance loss resulted from low-resolution PSs. 
 Numerical results have verified that the proposed DL-based E2E scheme can achieve significantly better performance than traditional schemes, especially in the case of limited channel estimation and feedback overhead. 

\end{spacing}
\begin{spacing}{1.38}

\end{spacing}
\end{document}